\documentclass{aa}
\usepackage{graphicx}
\usepackage{txfonts}
\usepackage{natbib}
\bibpunct{(}{)}{;}{a}{}{,}

	\newcommand{\be}{\begin{equation}}
	\newcommand{\ee}{\end{equation}}
	\newcommand{\ba}{\begin{array}}
	\newcommand{\ea}{\end{array}}
  \newcommand{\lir}{$\mbox L_{\mbox{\scriptsize IR}}$}
  \newcommand{\lirs}{$\mbox L_{\mbox{\scriptsize IR}}$~}
  \newcommand{\sfrir}{SFR$_{\mbox{\scriptsize IR} + \mbox{\scriptsize UV}}$}
  \newcommand{\sfrirs}{SFR$_{\mbox{\scriptsize IR} + \mbox{\scriptsize UV}}$~}
  \newcommand{\sfrfit}{SFR$_{\mbox{\scriptsize fit}}$}
  \newcommand{\sfrfits}{SFR$_{\mbox{\scriptsize fit}}$~}
  \newcommand{\sfruv}{SFR$_{1500}$}
  \newcommand{\sfruvs}{SFR$_{1500}$~}
  \newcommand{\mip}{24~$\mu$m}
  \newcommand{\mips}{24~$\mu$m~}

\begin{document}

\title{Star formation and mass assembly in high redshift galaxies}

   \author{P. Santini \inst{1,2}
   \and
   A. Fontana \inst{1}
   \and
   A. Grazian \inst{1}
   \and
   S. Salimbeni \inst{1,3}
   \and
   F. Fiore \inst{1}
   \and
   F. Fontanot \inst{4}
   \and
   K. Boutsia \inst{1}
   \and
   M. Castellano \inst{1,2}
   \and
   S. Cristiani  \inst{5}
   \and
   C. De Santis \inst{6,7}
   \and
   S. Gallozzi \inst{1}
   \and
   E. Giallongo \inst{1}
   \and
   N. Menci \inst{1}
   \and
   M. Nonino \inst{5}
   \and
   D. Paris \inst{1}
   \and
   L. Pentericci \inst{1}
   \and
   E. Vanzella  \inst{5}
   }

   \offprints{P. Santini, \email{santini@oa-roma.inaf.it}}

\institute{INAF - Osservatorio Astronomico di Roma, Via Frascati 33,
00040 Monteporzio (RM), Italy \and Dipartimento di Fisica,
Universit\`{a} di Roma ``La Sapienza'', P.le A. Moro 2, 00185 Roma,
Italy \and Department of Astronomy, University of Massachusetts, 710 North Pleasant Street, Amherst, MA 01003 \and MPIA Max-Planck-Institute f\"ur Astronomie, Koenigstuhl 17, 69117 Heidelberg, Germany \and INAF - Osservatorio Astronomico di Trieste, Via G.B.
Tiepolo 11, 34131 Trieste, Italy  \and Dip. di Fisica, Universit\`{a} Tor Vergata, Via della Ricerca
Scientifica 1, 00133 Roma, Italy \and INFN-Roma Tor Vergata, Via della Ricerca Scientifica 1, 00133 Roma, Italy}

   \date{Received .... ; accepted ....}
   \titlerunning{Star formation and mass assembly in high redshift galaxies}

   \abstract {} { The goal of this work is to infer the star formation
     properties and the mass assembly process of high redshift ($0.3
     \leq z < 2.5$) galaxies from their IR emission using the \mips
     band of MIPS-Spitzer.}  { We used an updated version of the
     GOODS-MUSIC catalog, which has multiwavelength coverage from 0.3 to \mips  and either spectroscopic or accurate photometric redshifts. We
     describe how the catalog has been extended by the addition of
     mid-IR fluxes derived from the MIPS \mips image.  We compared two
     different estimators of the star formation rate (SFR
     hereafter). One is the total infrared emission derived from
     \mip, estimated using both synthetic and empirical IR
     templates. The other one is a multiwavelength fit to the full
     galaxy SED, which automatically accounts for dust reddening and
     age--star formation activity degeneracies.  For both estimates,
     we computed the SFR density and the specific SFR.}  { We
       show that the two SFR indicators are roughly consistent, once
       the uncertainties involved are taken into account. However,
       they show a systematic trend, IR-based estimates exceeding the fit-based ones as the star formation rate increases.
     With this new catalog, we show that: \textit{a)} at $z>0.3$,
     the star formation rate is correlated well with stellar mass, and
     this relationship seems to steepen with redshift if one relies on 
     IR--based estimates of the SFR; \textit{b)} the contribution to the
     global SFRD by massive galaxies increases with redshift up to
     $\simeq 2.5$, more rapidly than for galaxies of lower mass, but
     appears to flatten at higher $z$; \textit{c)} despite this
     increase, the most important contributors to the SFRD at any $z$
     are galaxies of about, or immediately lower than, the characteristic stellar mass; \textit{d)} at $z\simeq 2$, massive galaxies are actively
     star-forming, with a median SFR $\simeq 300
     M_\odot$yr$^{-1}$. During this epoch, our targeted galaxies assemble a
     substantial part of their final stellar mass; \textit{e)} the specific SFR (SSFR) shows a clear bimodal distribution.}  { The analysis
     of the SFR density and the SSFR seems to support
     the \textit{downsizing} scenario, according to which high mass
     galaxies have formed their stars earlier and more rapidly than their
     low mass counterparts.  A comparison with renditions of
     theoretical simulations of galaxy formation and evolution indicates
     that these models follow the global increase in the SSFR with
     redshift and predict the existence of quiescent galaxies even at
     $z>1.5$.  However, the average SSFR is systematically
     underpredicted by all models considered.}


\maketitle

\section{Introduction}

Answering the basic questions about the birth, formation, mass build-up, and evolution of galaxies throughout cosmic time are some of the major goals of observational extragalactic astronomy.

In the past few years, this issue has been approached with two different methods. 
Many previous works measured a rapid evolution in the stellar mass density between $z \sim 1$ and $z \sim 3$ \citep{dickinson03,fontana03,fontana04,glazebrook04,drory04,fontana06,rudnick06,papovich06,yan06,pozzetti07} and demonstrated that a substantial fraction (30-50\%) of the stellar mass formed during this epoch. 
The differential evolution in the galaxy stellar mass function, according to which massive galaxies evolve rapidly up to $z\sim1.5$ and then more gradually until the present epoch, while less massive galaxies continue to evolve, implies that massive galaxies must have already formed by $z\sim1.5$. 
Several groups \citep[e.g.,][]{faber07,brown07} studied the evolution of massive galaxies at $z \lesssim 1$, and their migration from the blue cloud to the red sequence. 
We note that optical observations \citep{bell04,zucca06} suggest that the number of massive galaxies, as well as the stellar mass on the red sequence, has nearly doubled since $z \sim 1$, in qualitative agreement with the hierarchical merging scenario.

A parallel line of study has analyzed the rate at which galaxies form stars during different epochs and shown that they experience an extremely active phase in the same redshift range \citep[e.g.,][]{lilly96,madau96,steidel99,hopkins04,hopkins06,daddi07a}.  
Galaxies appear to form their stars following the so-called \textit{downsizing} scenario, in which the star formation shifts from high mass to low mass galaxies as redshift decreases. This picture was first introduced by \cite{cowie96}, who studied the evolution in the $K_s$ band luminosity function with redshift, and proposes that the most massive galaxies assemble their mass both earlier and more quickly than their lower mass counterparts, which, in contrast, continue to form stars until recent epochs. 
Later on, many other groups \citep[e.g.,][]{brinchmann00,fontana03,feulner05,perezgonzalez05,papovich06,damen09} derived confirmations of a \textit{downsizing} behaviour from the study of the specific star formation rate, defined to be the star formation rate per unit mass, at different redshifts. 
However, we note that deep radio observations \citep{dunne09} appear to conflict with this scenario. 
The \textit{downsizing} picture also appears to contradict the hierarchical growth scenario in which the most massive structures that we see today are produced by merging processes between smaller structures inside large-scale overdensities and collapsed when the Universe was far younger than today.

To reproduce this early formation of massive galaxies  \citep[see][]{thomas05} that are already ``red and dead'' at high $z$, theoretical models had to introduce very efficient processes of star formation and its suppression by means of active galactic nuclei and supernovae quenching of cooling flows \citep{menci06,kitzbichler07,bower06,croton06,nagamine06,monaco07},  gravitational heating \citep{khochfar08,johansson09}, or shock heating \citep{dekel06}. 
These models differ slightly in their predictions mainly because they adopt different processes to shut down the star formation. 

Both stellar masses and star formation rate estimates are affected by a number of uncertainties. The measure of the star formation rate (SFR) is especially difficult to handle. 
The high amount of energy produced  by newly born stars is emitted throughout the galaxy spectral energy distribution (SED), from X-rays to radio frequencies. Since we are of course unable to directly measure the total light emitted by young and massive stars, calibration factors and corrections are applied to estimate its value for any of these frequency ranges \citep{kennicutt98,bell03,calzetti08}. 
One of the most commonly used estimators is the UV rest-frame band, where young and massive stars emit most of their light. However, dust absorbs, reprocesses, and re-radiates UV photons at near-to-far IR wavelengths. Hence, the reliability of UV luminosity as a SFR tracer depends on large and uncertain corrections relying upon the dust properties, which are not yet clearly known  \citep{calzetti94,calzetti97,calzetti01}. 
Moreover, the UV-upturn at $\lambda$ shortward of 2500 $\AA$ \citep[e.g.,][]{han07}, especially in elliptical galaxies, can potentially bias the SFR estimate at very low redshift. 
Since the most intense star formation episodes are expected to occur in dusty regions, 
most of the power originating in star-forming (SF) galaxies is emitted in this wavelength range, and the dust emission peak is the dominant component of SF galaxies SEDs \citep{adelberger00,calzetti00}. 
Thus, a popular approach consists of adopting a conversion between the total emitted IR luminosity (\lirs hereafter) and a star formation rate estimation that is unaffected by dust obscuration \citep{kennicutt98}. 

The total infrared luminosity is generally estimated by comparing observed SEDs and synthetic templates, although  empirical conversions have sometimes been used \citep{takeuchi05,bavouzet07}. A variety of different libraries are used for this purpose \citep[e.g.,][and so on]{ce01,dh02,lagache03,siebenmorgen07}. A notable problem for the reliability of IR--based SFR tracers concerns obscured AGNs. In these objects the IR emission is generated by matter accretion onto a central black hole rather than dust heating by young stars. 

In this paper, we use the GOODS-MUSIC catalog to investigate properties of star-forming galaxies up to redshift 2.5 and infer the mass assembly process from their mid-IR emission. 
The paper is organized as follows. 
In Sect. 2, we recall the basic features of the GOODS-MUSIC dataset and explain the innovations concerning its latest version, and we explain how it has been updated with the addition of the \mips photometric band. 
We derive and compare star formation rates from IR-- and fit--based estimators in Sect. 3. In Sect. 4, we present a study of the mass assembly process in the high redshift Universe and a comparison with theoretical model predictions. 
We finally summarize our work and our conclusions in Sect. 5. 
In Appendix A, we describe in more detail how we convert mid-IR fluxes into total infrared luminosities and compare the different templates used, and in Appendix B we present the error analysis performed on the fit--based SFR estimates.

Throughout this work, unless  stated otherwise,  
we assume a \cite{salpeter55} initial mass function (IMF) and adopt the $\Lambda$-CDM concordance cosmological model (H$_0$ = 70 km/s/Mpc, $\Omega_{\small M}$ = 0.3 and $\Omega_{\Lambda}$ = 0.7).

\section{The data sample}

\subsection{The new GOODS-MUSIC sample}

We present and use an updated version of the multicolour
GOODS-MUSIC sample \citep[GOODS MUlticolour Southern Infrared Catalog;][]{grazian06},
extracted from the public data of the GOODS-South survey \citep{giavalisco04}. In the following, we shall refer to this version of the catalog as GOODS-MUSIC v2, to differentiate it from 
the former public version, which is named v1 hereafter. The new version is
also made publicly available \footnote{The catalog is available in electronic form
at the CDS via anonymous ftp to cdsarc.u-strasbg.fr (130.79.128.5)
or via http://cdsweb.u-strasbg.fr/cgi-bin/qcat?J/A+A/. It is also possible to download the catalog at the WEB site http://lbc.mporzio.astro.it/goods.}.

The 15-bands multiwavelength coverage ranges from 0.35 to \mip,
as a result of the combination of images from different
instruments (2.2ESO, VLT-VIMOS, ACS-HST, VLT-ISAAC, Spitzer-IRAC,
Spitzer-MIPS). 
The catalog covers an area of $\sim$ 143.2 arcmin$^2$
located in the Chandra Deep Field South and consists of 15\,208 sources. 
After culling Galactic stars, it contains 14\,999 objects selected in either the $z$ band or the $K_s$ band or at 4.5 $\mu$m.

The whole catalog has been cross-correlated with spectroscopic
catalogs available to date, and a spectroscopic redshift has been
assigned to $\sim$12 \% of all sources.  For all other objects, we have
computed well-calibrated photometric redshifts using a standard $\chi^2$ 
minimization technique for a large set of synthetic spectral templates.

The previous version of the catalog and procedures adopted to determine the photometric
redshifts and physical properties of each object were described at
length in \cite{grazian06} and \cite{fontana06}.  With respect to the
previous catalog, we have performed a set of improvements to the
optical--near-IR data, the major ones being:

\begin{itemize}
\item {In addition to objects selected in the ACS $z$ and in the ISAAC
    $K_s$ bands, we have also included objects selected from the IRAC 4.5
    $\mu$m image, hence including sources detected at 4.5
    $\mu$m but very faint or undetected even in $K_s$ band. A full
    description of these objects is beyond the scope of the present
    paper and will be presented elsewhere.}
\item {We have revised the photometry in the four IRAC bands
    using an updated version of the PSF-matching kernels, as released by the Spitzer Science Center. As a consequence, we also adopted a larger kernel, to fully account for the large tails of the IRAC PSFs.}
\item {We have adopted a revised procedure for estimating the
    background in the IRAC images. 
    Using the objects' positions and IRAC fluxes from the GOODS-MUSIC v1 catalog, we have created realistic simulated images in the four IRAC bands by smoothing sources to the nominal IRAC PSFs. An accurate background estimation has been performed by subtracting these simulated images from the original ones and by linearly interpolating  the residual emission.    
    Since the average value of the
    apparent IRAC background is negative, this has led to an increase
    in the adopted background, with respect to the GOODS-MUSIC v1
    version. }
\item {These two changes have modified the IRAC
    photometry. Because of the new kernels, the brightest objects have a higher flux and 
    a typical offset in magnitude of 0.23, 0.14, 0.22, 0.35, respectively in the 3.6, 4.5, 5.8, and 8 $\mu$m  images, with respect to the GOODS-MUSIC v1 catalog, which agrees with the analysis of \cite{wuyts08}. This effect is largely
    mitigated for fainter galaxies, since the higher background now
    adopted has led to an underestimate of their fluxes.}
\item { Overall, the revised IRAC
    photometry has  a modest impact on the estimate of photometric redshifts,
    since $ \langle z_{phot\ v2} - z_{phot\ v1} \rangle \sim 0.01 \pm 0.16$  for the whole sample and $\sim 0.01 \pm 0.03$  when a 3$\sigma$--clipping analysis is performed (see also \cite{wuyts08}). 
}
\item {A more informative test of the photometric
    redshift accuracy comes from the enlargement of the sample of galaxies
    with spectroscopic redshifts, which we obtained by adding new
    spectra from public surveys  \citep{vanzella08,popesso09}. In addition, we have also had access to the spectra of the GMASS survey \citep{cimatti08}, prior to
    their publication. The final sample now includes 1888  
    galaxies, three times larger than the spectroscopic sample in
    \cite{grazian06}. The additional spectra are mostly relative to
    galaxies that are both fainter and at higher redshift than in the
    original sample. Without significant refinements in the adopted
    templates, we then find that the absolute scatter $|\Delta z| =
    |z_{spe}-z_{phot}|/(1+z_{spe})$  has a slightly larger average value. Quantitatively, the average absolute scatter is now $\langle|\Delta z|\rangle =0.06$, instead of 0.045 obtained for the GOODS-MUSIC v1 catalog. However, when only the brightest galaxies are considered, we find comparable values with respect to \cite{grazian06} ($\langle|\Delta z|\rangle =0.043$). We have verified that this is due to an increased number of outliers, as shown by a 3$\sigma$--clipping analysis, which provides $\langle|\Delta z|\rangle = 0.027$ and $0.032$ for the complete datasets of v1 and v2 catalogs, respectively.}
\item {We have removed Galactic stars and performed a more careful selection of the galaxy sample to identify AGN sources. For the latter, we have
    first removed all objects whose spectra show AGN features. 
    Then, we have cross-correlated our catalog with the
    X-ray catalog of \cite{brusa09}, and removed all X-ray detected sources whose
    flux is dominated by an unresolved central source. These sources
    typically have spectra classified as narrow-line AGNs. The optical
    morphologies of all remaining X-ray sources do not show a
    dominating central point like source, and -- where available -- have
    typically spectra classified as emission-line star-forming
    galaxies. These objects have been retained in our galaxy sample. 
    }
\end{itemize}

The major new ingredient of this new version of the GOODS-MUSIC
catalog, however, is the inclusion of the \mips photometry for all
galaxies in the sample, which is the main focus of the present
paper. We describe the adopted procedures and results in the
following.

\subsection{MIPS \mips catalog} \label{sec:mips24}

We have extended the GOODS-MUSIC catalog with the addition of the
mid-IR fluxes derived from the public \mips image of the Multiband
Imager Photometer for Spitzer \citep[MIPS,][]{rieke04} onboard the Spitzer
Space Telescope. As for the IRAC images, given the very
large PSF of this image ($\sim 5.2$ arcsec), to properly detect and
de-blend objects we had to employ a PSF-matching
technique, which is performed by the software ConvPhot
\citep{desantis07}. This algorithm measures colours from two images of
different qualities by exploiting the spatial and morphological information
contained in the higher resolution image. 

When applying ConvPhot to
our case, each object was extracted from the high resolution $z$ band
ACS-HST (PSF $\sim 0.12^{''}$) image, which was used as a {\it prior} to
extract the objects' positions, filtered with a convolution
kernel, and finally scaled by a $\chi^2$ minimization over
all the image pixels to match the intensity in the MIPS image. 
To fully use the positional information of the ACS images, and maintain
consistency with the $z$--selected sample, we used the {\it z} band ACS-HST data as a {\it prior},
augmented by artificial objects placed where galaxies detected only in
$K_s$ or 4.5 $\mu$m were located. 
In our case, the MIPS-Spitzer and ACS-HST images have pixel scales of
1.2 and 0.03 arcsec/pixel, which means that it is impracticable to use 
ConvPhot even with fast workstations.  To make the computation
feasible, we rebinned the ACS detection image by a factor 8 $\times$ 8
(0.24 arcsec/pixel). 
In regions where the crowding of the $z$-detected sources was significant,
the fit may become unconstrained or degenerate because of the large size
of the MIPS PSF. To prevent this, we placed an additional constraint on
the fitted fluxes that must be non-negative.  For the objects whose
flux is forced to be zero, we provide an upper limit derived from the
analysis of the mean rms in the object area.  

The behaviour of ConvPhot in these extreme applications was tested
with several simulations, which were described by
\cite{desantis07}. These tests indicated that the estimated magnitudes  are
 biased by neither the different qualities of the two images nor the
undersampling of the high resolution image after rebinning.

Finally, from visual inspection of sources with unusual colours, we removed $\sim 30$ objects from the catalog  whose flux was incorrectly assigned. By examination of the residuals, we also verified that there is no significantly bright source in the MIPS image apart from the ones considered.
We detected 3313 ($\sim 22\%$ of total) detected objects and 11 841 ($\sim 78\%$)
1$\sigma$ upper limits.

Despite the validation tests providing satisfying results, we emphasize that the
intrinsic limitations due to the poor resolution of the MIPS image
cannot be completely overcome, in particular for sources 
blended even in the \textit{detection} image, i.e., sources whose
profiles overlap in the ACS {\it z} image. In this case, the separation
between the two -- or more -- objects can become even smaller than the
positional accuracy of the MIPS images, and the association between
the $z$-detected and the MIPS sources relies on the accuracy of the
astrometric solution.  To check for possible
misidentifications, we associated each source with a flag to
quantify the number of possible contaminants, at different distances. 
This flag was attached to the public catalog and we caution the user to check for 
possible systematic errors. 
The results that we shall present in the following are insensitive to the inclusion
of the most blended sources, which  have therefore not been removed.  

\begin{figure}[t]
\resizebox{\hsize}{!}{\includegraphics{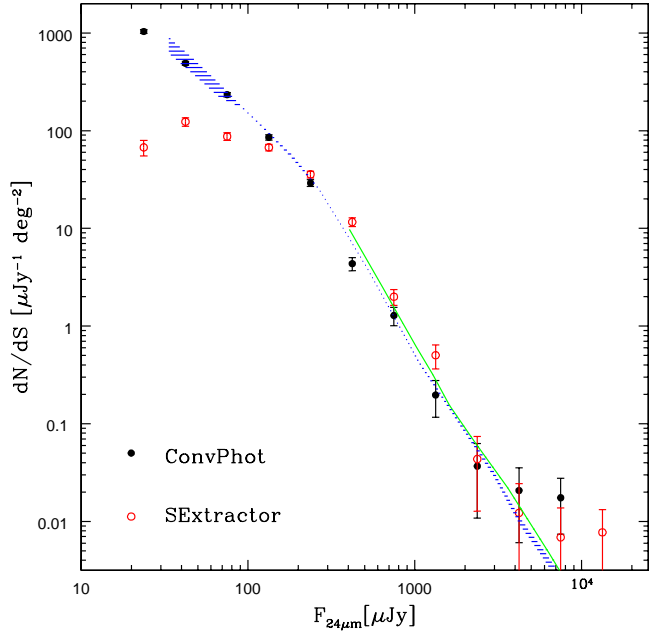}}
\caption{\mips flux number counts measured by ConvPhot (black filled
  circles) and SExtractor (red empty circles). Error bars have been computed as
  the square root of the number of objects in each bin. We compared our results
 to \cite{papovich04} and \cite{shupe08} \mips counts (blue shaded region and green line, respectively). }
\label{fig:lognlogs}
\end{figure}

It may be interesting to observe the effect of this procedure on the 
\mips number counts, shown in Fig. \ref{fig:lognlogs}.  We present
both the counts derived by ConvPhot as well as those obtained by a
SExtractor catalog, which was produced by adopting corrected
aperture magnitudes at 6 arcsec.
The blue shaded region shows the \mips counts from
\cite{papovich04}. They counted sources in five different fields
(CDFS among them) using DAOPHOT software.  The green
solid line (error bars are not large enough to be seen) represents
\cite{shupe08} counts in the SWIRE field performed by SExtractor.
ConvPhot (black dots) provides consistent results with both the previous work of
other authors to the lowest fluxes and our SExtractor catalog (red circles)  at $F\gtrsim 100\ \mu$Jy.

The agreement at bright fluxes confirms the results of the simulations and validation
tests presented in \cite{desantis07} and shows that the fluxes
estimated by ConvPhot agree with those estimated by an independent
detection with SExtractor for sources that are not severely
blended. For the blended fraction of the objects, however, the
fluxes estimated by ConvPhot can be slightly lower, since part of the flux is identified with
the fainter contaminants, which are not detected in the \mips image
alone. 

It is more interesting to consider the behaviour at faint fluxes, where
the a priori knowledge of the object position, because of the use of the
$z$, $K_s$ and 4.5 $\mu$m images for the detection, allows us to obtain
flux estimate at much fainter limits, reducing the effects of blending
and confusion.  As expected, indeed, SExtractor counts decline at
$\sim$100 $\mu$Jy, where the confusion limit prevents the detection of
fainter sources, while ConvPhot allows us to complete our detection to even deeper limits.  ConvPhot number counts present a double slope, with a break point located at $\sim$100~$\mu$Jy, which we consider to be an intrinsic property of the sample. The slope and the
normalization at the faint end agree with the estimates of
\cite{papovich04}, who carefully computed a correction for the
incompleteness due to poor resolution at faint limits.
\cite{papovich04} and subsequent papers \citep[e.g.,][]{lefloch05,perezgonzalez05,papovich06,marcillac06,bell07} estimated that the source
detection in MIPS-CDFS is 80\% complete at 83 $\mu$Jy or so, which typically
corresponds to $S/N \sim 28-30$ in our fitting procedure.  For
consistency with these works, we therefore distinguish between
objects with fluxes above this limit and those detected at lower
$S/N$, to fluxes as low as $\sim$20 $\mu$Jy, which corresponds to our flux
count limit. The median $S/N$ at this flux limit is $\sim 6-7$,
although a tail is present at lower values of $S/N$ that is caused by 
source blending. Nevertheless, this tail includes only a small number
of objects.

Finally, we cross-correlated our \mips catalog with both the FIREWORKS
catalog \citep{wuyts08} and the one released by the GOODS
Team\footnote{GOODS-South MIPS 24 micron source list v0.91 from GOODS
  data release (DR3).} \citep{chary06}, who
adopted a similar source extraction technique. The overall agreement
was good with a small offset in both cases
($\langle F_{24\mu m\ GOODS-MUSIC}/F_{24\mu m\ FIREWORKS} \rangle \sim1.2$ and
$\langle F_{24\mu m\ GOODS-MUSIC}/F_{24\mu m\ DR3}\rangle \sim 0.91$).

\subsection{The data selection}

As pointed out above, we removed from our catalog Galactic stars and both spectroscopic or X-ray detected AGN sources. Moreover, we only consider the redshift range 0.3 -- 2.5. 
From this sample, we consider the following two subsamples: 
\begin{itemize}
\item{the purely $K_s$--selected sample, subsample A, consists of 2602 galaxies. Of these, 981 (376) are \mips detections with F$_{24 \mu m} >$ 20 (83) $\mu$Jy  and 1621 have been assigned an upper limit;}
\item{subsample B was created by performing the following cuts: $z<26$ or $K_s<23.5$ or $m_{4.5}<23.2$. It includes 7909 galaxies, of which 1165 (413) are \mips detections with F$_{24 \mu  m} >$ 20 (83) $\mu$Jy and 6744 are upper limits. 
}
\end{itemize}

\section{Comparison between SFR indicators}

In this section, we present different methods to estimate the star formation rate and compare the results obtained from the different indicators.

\subsection{SFR estimators: IR--, fit-- and UV--based SFR} \label{sec:sfrestimate}

Based on the assumption that most of the photons originating in newly formed
stars are absorbed and re-emitted by dust, the mid-IR emission is in
principle the most sensitive tracer of the star formation rate. In
addition, a small fraction of unabsorbed photons will be 
detected at UV wavelengths. A widely used SFR indicator is therefore based on a
combination of IR and UV luminosity, which supplies complementary
knowledge of the star formation process \citep{iglesiasparamo06,calzetti07}.  For \mips detected
sources, we estimated the instantaneous SFR using the same calibration
as \cite{papovich07} and \cite {bell05}: 
\be
\mbox{SFR}_{\mbox{\scriptsize IR} + \mbox{\scriptsize UV}}/\mbox{M}_\odot \mbox{yr}^{-1} = 1.8 \times 10^{-10} \times \mbox{L}_{bol}/\mbox{L}_\odot \\
\label{eq:sfr}
\ee \be \mbox{L}_{bol} = (2.2 \times \mbox{L}_{\mbox{\scriptsize UV}}
+ \mbox{L}_{\mbox{\scriptsize IR}}) \ee

We computed \lirs by fitting \mips emission to \cite{dh02} (DH hereafter) synthetic templates, which are widely adopted in the literature. In Appendix A, we compare these values with the corresponding \lirs predicted by different model libraries \citep{ce01,polletta07}.  
We include the rest-frame UV luminosity, uncorrected for extinction, derived from the SED fitting technique, L$_{\mbox{\scriptsize UV}} = 1.5 \times \mbox{L}_{2700 \mbox{\scriptsize \AA}}$; although often negligible, this can account for the contribution from young unobscured stars. 

Following \cite{papovich07}, we then applied a lowering correction to the estimate obtained from Eq. \ref{eq:sfr}. 
They found that the \mips flux, fitted with the same DH library, overestimates the SFR with respect to the case where longer wavelengths (70 and 160 $\mu$m MIPS bands) are considered as
well, and they corrected the trend using an empirical second-order polynomial. In Appendix A we show further confirmation of the need to apply this correction: for bright sources, \lirs estimated by synthetic models has values of up to a factor of 10 higher  
than \lirs predicted by the empirical library of \cite{polletta07}. Similar results were also published by \cite{bavouzet07} and \cite{rieke09}. 
In the following, we refer to the estimate in Eq. \ref{eq:sfr} as \sfrir. 

A complementary approach to estimating the star formation rate, as well as
other galaxy physical properties (e.g., mass, age, dust extinction), is the SED fitting. A grid of spectral templates is computed from standard spectral synthesis models, and the expected magnitudes in our filter set are calculated. The derived template library is compared with
the available photometry and the best-fit model template was adopted according to a $\chi^2$ minimization. During the fitting process, the redshift is fixed to its spectroscopic or photometric value. 
The physical parameters associated with each galaxy are obtained from the best-fit template up to 5.5 $\mu$m rest-frame. 
This analysis assumes that the overall galaxy SED can be represented as a
purely stellar SED, extincted by a single attenuation law, and that
the relevant $E(B-V)$ and basic stellar parameters (mostly age and
star formation history, but also metallicity) can be simultaneously
recovered with a multiwavelength fit.  We note that
parameter degeneracies cannot be completely removed, especially at
high redshift. Previous studies \citep{papovich01,shapley01,shapley05}
demonstrated that, while stellar masses are well determined, the SED
fitting procedure does not strongly constrain star formation histories
at high redshifts, where the uncertainties become larger due to the SFR--age--metallicity degeneracies. 
For this reason, the uncertainties associated with the SFR values estimated from the SED fitting are larger than those associated with the IR tracer.

In our analysis, we estimated star formation rates (along with stellar masses) using \cite{bc03}
synthetic models, fitting the whole 14 bands of photometry (from the $U$
band to 8 $\mu$m).  We parameterize the star formation histories with a
variety of exponentially declining laws (of timescales $\tau$
ranging from 0.1 to 15 Gyr), metallicities (from $\mbox Z = 0.02~
\mbox{Z}_\odot$ to $ \mbox Z = 2.5~ \mbox{Z}_\odot$) and dust
extinctions ($0 < \mbox{E(B-V)} < 1.1$, with a Calzetti or Small
Magellanic extinction curve). Details are given in Table 1 of
\cite{fontana04}, in \cite{fontana06} and in
\cite{grazian06,grazian07}. With respect to these
recipes, the only difference in our method is the adoption of a minimum age of 0.1 Gyr. Below this
value, the relation between UV luminosity and star formation rate
changes rapidly with the age of the stellar population,
leading to very high values of inferred SFRs. We are aware that exponential star formation histories may not be the correct choice in some cases. However,  modeling in detail the star formation history of our galaxies is beyond the scope of the present paper, whose aim is to compare the star formation rates derived from the IR emission with those based on the widely used  SED fitting procedures. 
In the following, we refer to this SFR estimation as \sfrfit.

We have also fitted our data using the \cite{maraston05} (MR05) and Charlot \& Bruzual \citep[in prep., see][]{bruzual07a,bruzual07b} (CB07), including an improved TP-AGB stars treatment. The stellar mass estimates inferred using these new models are presented in \cite{salimbeni09} and are approximately 0.2 dex lower than those computed with \cite{bc03} models. However, the extrapolated SFR values lead to a significant overestimate (especially in the case of \cite{maraston05} models)  of the SFR density that we present in Sect. \ref{sec:sfrd}. We suspect that part of this discrepancy is caused by the peculiar shapes in the near-IR side of the spectrum, which probably cause the worse $\chi^2$ than measured from the comparison with \cite{bc03} models. Since a robust comparison among different stellar population models is beyond the scope of this paper, and pending further tests on the new MR05 and CB07 models, we continue to adopt the widely used \cite{bc03} template library and refer to possible future work for more details on this point.

At $z>1.5$, it is possible to obtain an independent estimate of the
SFR using the observed $L_{1500}$ and the observed slope of the UV
continuum to estimate the $E(B-V)$, rather than the multiwavelength
fit. 
In the following, we use the conversions adopted by \cite{daddi04},
culling a sample of BzK--SF galaxies from our catalog and deriving the
relevant SFR using the \cite{daddi04} scaling relations
$SFR_{UV}/M_\odot yr^{-1} = L_{1500}/L_0$, with $L_0=8.85 \times10^{27} erg\ s^{-1} Hz^{-1}$, 
and $A_{1500} = 10 \times E(B-V)$, where $E(B-V) = 0.25(B-z+0.1)_{AB}$.  
We refer to this SFR estimation as \sfruv.

For clarity, SFRs lower than 0.01 $M_\odot yr^{-1}$ are assigned a value 0.01 $M_\odot yr^{-1}$ in the following analysis.  

\subsection{Comparison between IR-- and fit--based  tracers} \label{sec:sfrcomp}

\begin{figure}[t]
\resizebox{\hsize}{!}{\includegraphics{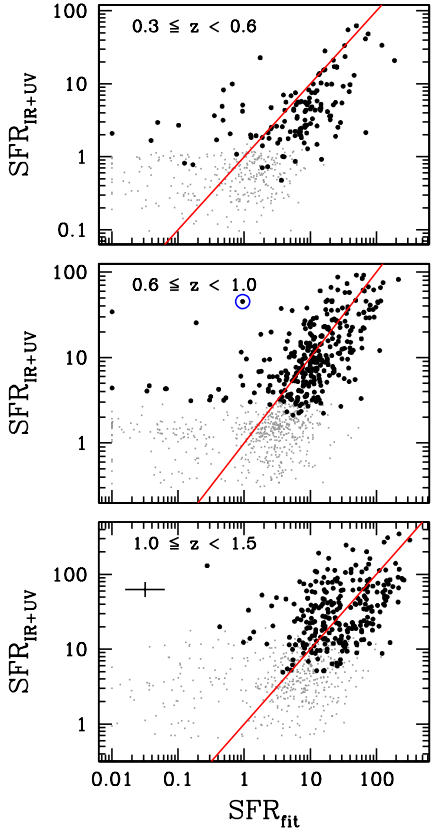}}
\caption{Relation between \sfrirs and \sfrfits in three redshift bins
  for the whole $K_s$--selected sample. Filled black dots are for the
  $F_{24\mu m} > 20\ \mu Jy$ subsample; small gray dots refer to galaxies
  undetected at \mip, and must be regarded as upper limits.  The blue circle identifies an obscured AGN candidate, selected according to a similar technique as the one presented in \cite{fiore08} (see text).  The red line defines the locus \sfrirs $=$ \sfrfit. The typical uncertainties associated with both estimates are shown in the lower panel.  The intrinsic parameter degeneracy involved with the SED fitting procedure is responsible for the larger uncertainties associated with \sfrfit. Units are $M_\odot yr^{-1}$. }
	\label{fig:sfr0}
\end{figure}

We now discuss the consistency between the different SFR estimates. In this section, we
use the $K_s$--selected sample (subsample A) to ensure a proper
sampling of the full SED, e.g., to ensure that all the bands, or most of them, are available for the fitting procedure.  

We start from the redshift range $0.3-1.5$,
plotting in Fig. \ref{fig:sfr0} the comparison between \sfrirs and
\sfrfit. We include both the \mips detected and undetected galaxies, for which only upper limits to \sfrirs can be 
obtained.  These upper limits are  misleading for the interpretation of the figure: since the SED fitting can reach lower nominal values for the given SFR, the scatter in \sfrfits seems to be larger than that in \sfrir.

Given the many uncertainties involved in both estimators, the overall
consistency between them appears reassuring. Apart from offsets and
other systematics, which we discuss below, the majority of
galaxies are assigned a consistent SFR, and the number of severe
inconsistencies is small.  These inconsistencies may have two
different origins.  

On the one hand, galaxies with \sfrfits much
higher than \sfrirs can in principle be inferred from incorrect fitting of
red galaxies. For these objects, the SED fitting could erroneously
assign a large amount of dust to an otherwise dust-free, passively
evolving population. 
Despite the relatively large number of passively
evolving galaxies at $z<1.5$, the number of these misidentifications is
very small at $z<1$, and low even at $z=1-1.5$.

\begin{figure*}[t]
\begin{tabular}{rl}
\resizebox{\hsize}{!}{
\includegraphics{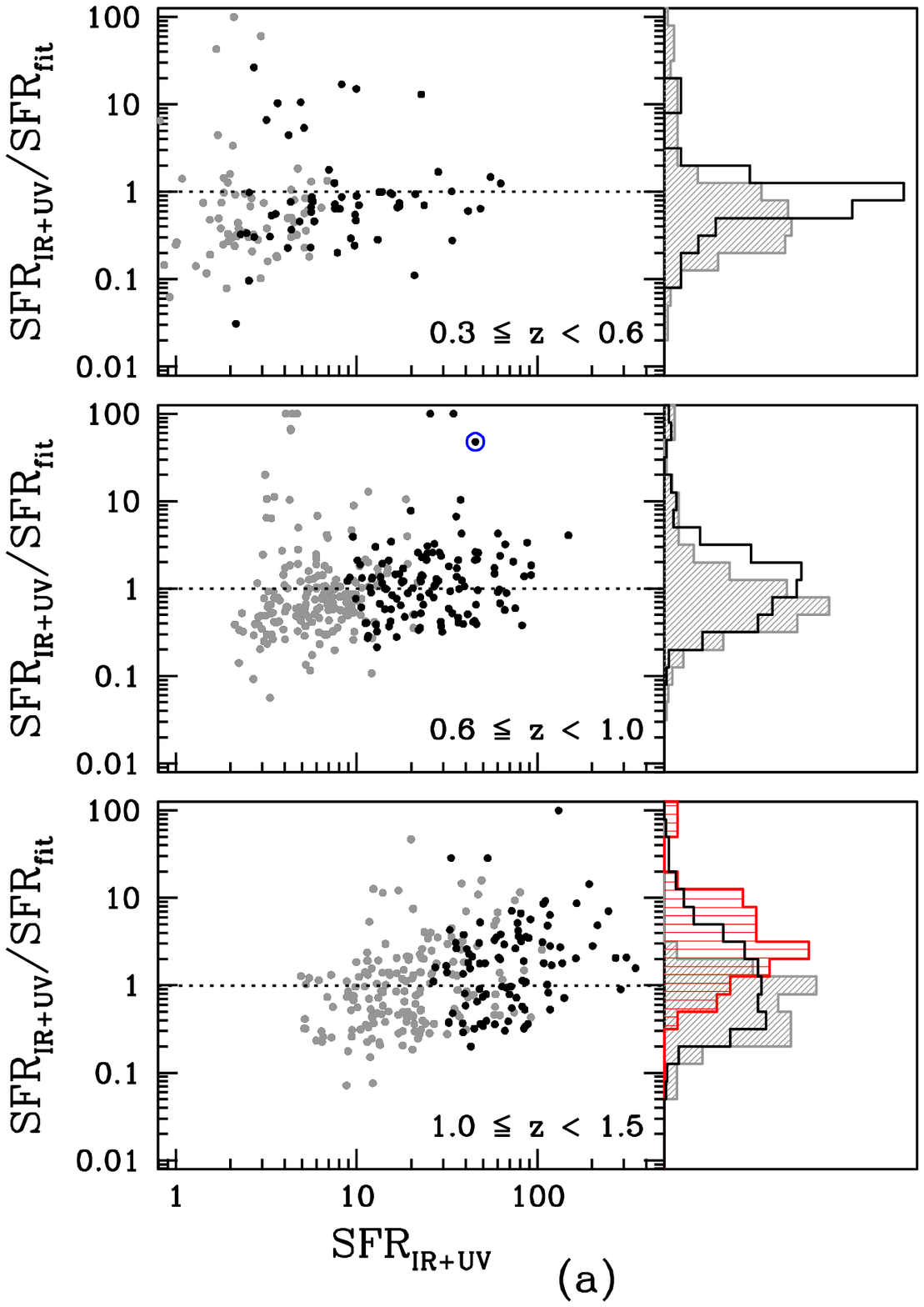}
\includegraphics{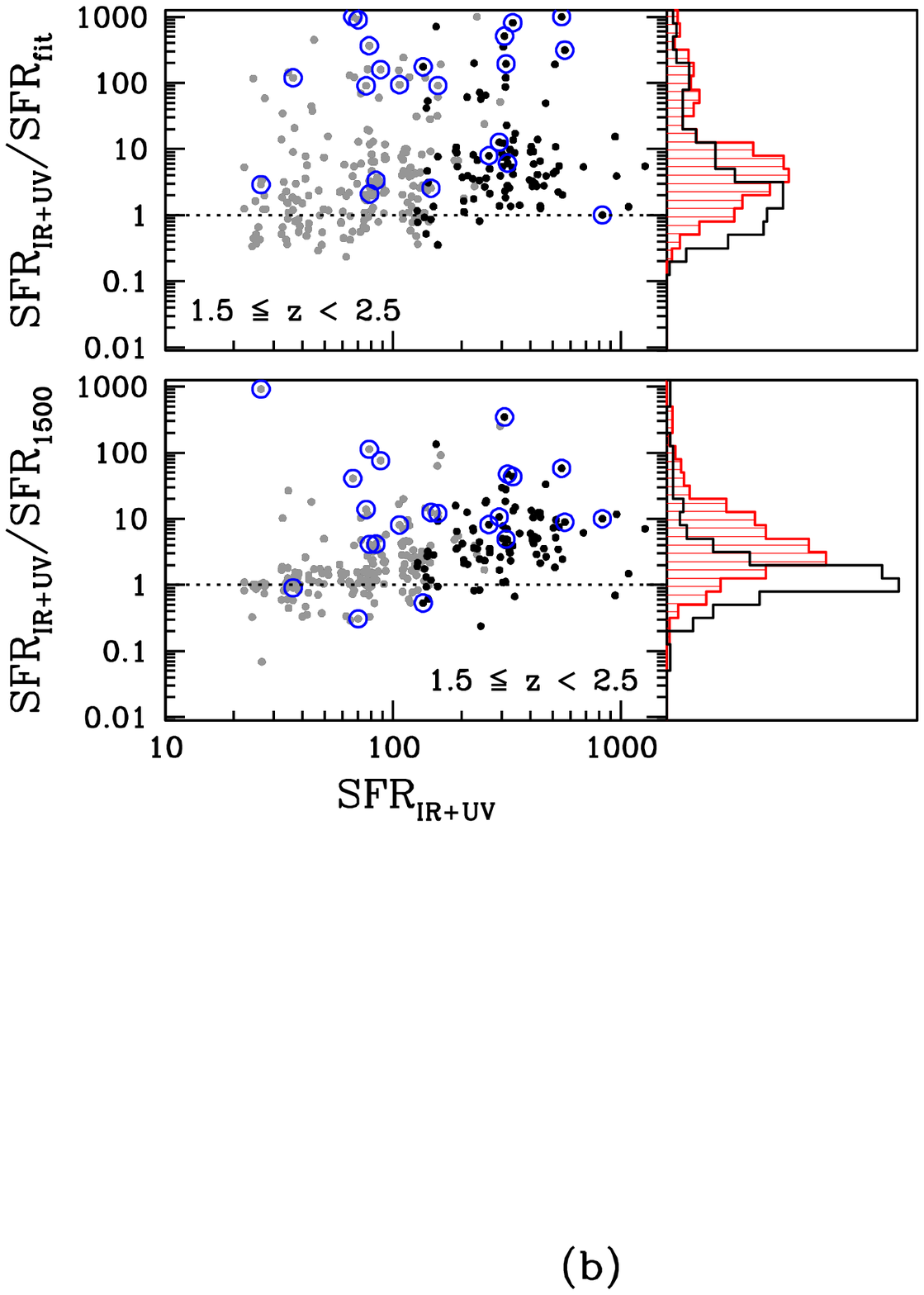}
}
\end{tabular}
\caption{Left panels in (a) and left top panel in (b): relation between \sfrir /\sfrfits ratio and \sfrirs in the different redshift bins for the $F_{24\mu m} > 83\ \mu$Jy (black dots) and $20 < F_{24\mu m}[\mu \mbox{Jy}] \leq 83$ detected galaxies (gray dots); blue circles identify obscured AGN candidates, selected according to a similar technique as the one presented in \cite{fiore08} (see text). Left middle panel in (b): relation between \sfrir /\sfruvs ratio and \sfrirs in the redshift bin 1.5--2.5 for the subsample of BzK-SF galaxies; symbols are as in the other panels. Right panels in (a) and (b): black plain, gray (diagonally) shaded and red (horizontally) shaded histograms show \sfrir /\sfrfits  (\sfrir / \sfruvs in the middle panel in (b)) values respectively for 10$<$\sfrir$[M_\odot yr^{-1}]<100$, \sfrir$[M_\odot yr^{-1}]<10$ and \sfrir$[M_\odot yr^{-1}]>100$ samples. Units are $M_\odot yr^{-1}$. }
   \label{fig:sfrratio}
\end{figure*}

On the other hand, galaxies with \sfrirs $>>$ \sfrfits can be obtained
either when the opposite misidentification occurs (e.g., for dusty
star-forming galaxies fitted with a passively evolving SED) or, more
interestingly, when the mid-IR emission is due to additional
processes, not observable in the UV/optical regime. Typical cases are
AGN emission or additional star formation activity that is completely dust
enshrouded. At $z<1.5$, these objects are again very rare in
our sample. In particular, the SFR of the  few galaxies with
\sfrirs$\simeq 2-6\ M_\odot yr^{-1}$ and \sfrfits$<1M_\odot yr^{-1}$ at
$z<1$ is probably overestimated, because of the incorrect
application of a star-forming template to a more quiescent galaxy
(see Fig. \ref{fig:lirpol}).

For a more detailed discussion of the robustness of the fit--based SFR estimates, the related error analysis and $\chi^2$ contours, we refer the reader to Appendix B.

The existing systematic trends can be appreciated by examining Fig.
\ref{fig:sfrratio} (a), where we plot the \sfrir/\sfrfits ratio for the \mips
detected sample.  First, we note that the scatter in the
\sfrir/\sfrfits distribution widens with redshift. This is probably due
to a combination of effects: as we move to high redshifts,
galaxies become intrinsically fainter and the rest--frame spectral
coverage becomes narrower, making the SED fitting more uncertain;
likewise, \mips based indicators could become more uncertain as the
filter moves away from the rest--frame mid-IR and approaches the PAH
region.

More interestingly, a correlation between the \sfrir/\sfrfits ratio and
\sfrirs can be observed. If we focus our attention on galaxies
with star formation rates of between 10 and 100 $M_\odot yr^{-1}$, the \sfrir/\sfrfits
distribution (black histograms) is centred on unity and fairly
symmetric. In contrast, galaxies with milder activity (gray shaded
histograms) have lower \sfrir/\sfrfits ratios, and we find a first hint of
highly star-forming galaxies (red horizontally shaded histogram),
i.e., objects with star formation rates higher than 100 $M_\odot yr^{-1}$, that have higher
\sfrir/\sfrfits ratios.

The same systematic effects are seen at higher redshift. The upper
panel of Fig. \ref{fig:sfrratio} (b) shows the ratio \sfrir/\sfrfits in the
$1.5-2.5$ range. It is immediately clear that the spread is larger
than at lower redshifts, and that a large number of objects with
\sfrir/\sfrfits $>>$ 1 is observed.  The observed spread is not
surprising, since a similar disagreement, of up to two orders of
magnitude in single galaxies at comparable redshifts, has already been
noticed by the similar analysis of \cite{papovich06}.  At these
redshifts, the faintness of the galaxies and the large
$k$--corrections in the mid-IR are clearly even more effective in
increasing the noise in both estimates.  
The factor of $\sim$2 shift in the distribution of galaxies with SFRs of between 10 and 100 $M_\odot yr^{-1}$ is consistent with the uncertainty associated with the spectral library used to compute \lirs (see Appendix A).

It is therefore interesting to compare these estimates with the pure
UV--based \sfruv, described above.  The relation between \sfrirs and
\sfrir/\sfruvs ratio is shown in the middle panel of Fig.
\ref{fig:sfrratio} (b). Following \cite{daddi04}, we applied this recipe only to
the sample of BzK--SF in the redshift range of interest. 
The scatter
is significantly lower than in the upper panel, but the same
trend for exhibiting an IR excess still appears for highly star-forming
objects.

To some extent, the excess in the mid-IR derived star formation rate is
most probably caused by the presence of highly obscured AGNs.
\cite{daddi07b} assumed that all the mid-IR excess
objects drawn from the BzK-SF sample (i.e., all objects in the lower panel of Fig.
\ref{fig:sfrratio} (b) with \sfrir/\sfruvs $>3$) are powered by
obscured AGNs. Following a different approach, \cite{fiore08}
identified a population of highly obscured AGNs candidates in the
entire $K_s$--selected sample, selected on the basis of their very red spectrum
($F_{24\mu m}/F(R)>1000$ and $R-K_s>4.5$) \citep[see also][]{dey08}. 
To use the fluxes directly measured in our catalog, we selected the very same objects following the criterium  $F_{24\mu m}/F(I)>1000$ and $I-K_s>4.5$, which we checked to be consistent with that used in \cite{fiore08}. 
These objects are shown as blue open
circles in Figs. \ref{fig:sfr0} and \ref{fig:sfrratio}.

The presence of highly obscured AGNs could also produce the observed
trend in the \sfrir/\sfrfits ratio, since they are expected to exist
inside high mass galaxies, which are on average highly star-forming.
A correlation between the fraction of obscured AGNs and the
stellar mass was indeed shown in \cite{daddi07b}.

However, we find that the trend in the \sfrir/\sfrfits ratio 
also extends to lower star formation rates (of average values lower than
unity) and at lower redshifts. It is therefore possible that it
reflects a change in the intrinsic physical properties of
star-forming galaxies.  A possible explanation is related to
metallicity effects. Galaxies with sub--solar metallicities have lower
mid-IR emission \citep[at least at $8 \mu$m,][]{calzetti07} and higher
UV luminosity than solar ones, for a given level of SFR. The observed
trend can therefore be related to a metallicity trend, which is natural
to expect given the observed mass--metallicity relation from low to
high redshifts \citep[][and references therein]{maiolino08}. 
Unfortunately, a direct check of the statement above is not feasible. Reliable metallicities cannot be inferred from broad-band SED fitting, and high resolution spectroscopy is necessary to properly distinguish between SEDs characterized by different lines and hence metallicities.

Alternatively, the observed trend can be taken as evidence of a
failure in the assumption that a single attenuation law can adequately
model the output from a star-forming galaxy. 

\section{Mass assembly and \textit{downsizing}}

We now discuss the star formation properties of our galaxy sample as
a function of redshift and stellar mass.  
With respect to other surveys, our sample has the distinctive
advantage of being selected by a multiwavelength approach. In
this section, we use the subsample B. 
The $K_s$ and $m_{4.5}$ cuts ensure a proper sampling of
highly absorbed star-forming galaxies, and hence probably a complete
census of all galaxies with high SFR. On the other hand, the deep $z$--selected sample
contains the fainter and bluer galaxies of both low levels of dust
extinction and low star formation rate.

In this catalog, we derive the SFR, which we refer to as IR--based, using the \sfrirs estimates
to all objects with $F_{24\mu m} \geq 20\ \mu$Jy, and the \sfrfits to
all fainter objects. We recall that the \sfrirs star formation rates
are derived from the mid-IR emission, computed with the DH synthetic
models, and by adopting the lowering correction of \cite{papovich07} for high SFRs. 
This technique was widely adopted in the literature, and we adopt it as a baseline. 
At the same time, we 
mention how the results would be changed by using \sfrfits for all the objects. 
We  use the stellar masses estimates derived from the SED
fitting analysis  described above. 

Finally, to avoid  bias in the IR--based SFR estimates, we remove the obscured AGN candidates identified by \cite{fiore08} from our analysis. We are aware that the removal of this population will cause a reduction in the star formation rate measured on average at high masses. AGNs are indeed known to reside preferentially in high mass sources \citep{best05,daddi07b,brusa09}; although a high fraction of IR emission is understood to originate from accretion processes, some fraction of it probably originates from star formation.

\subsection{The stellar mass--SFR relation}

\begin{figure}[!t]
\resizebox{\hsize}{!}{\includegraphics[angle=0]{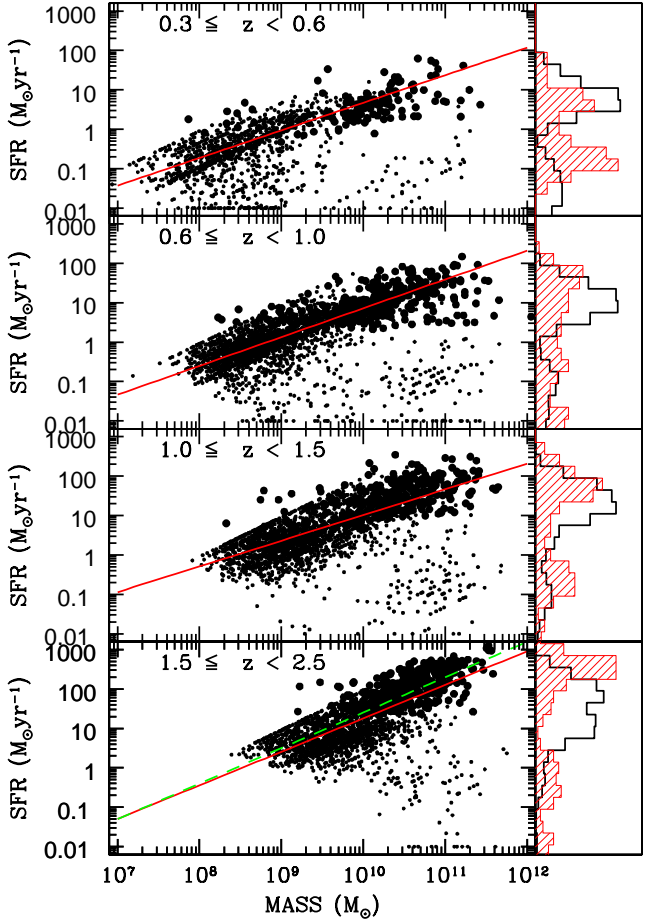}}
\caption{Left panels: relation between the star formation rate and the stellar mass
  of the GOODS--MUSIC galaxies at different redshifts; large dots
  represent the galaxies with SFR derived from the \mips emission, while small ones are \mips undetected galaxies, with SFR derived from the SED fitting analysis; red solid lines show the $2\sigma$--clipped least square fit described in the text; green dashed line in the highest redshift bin represents the correlation found by \cite{daddi07a}.  Right panels: SFR distribution in two mass bins, $10^{10}$ -- $10^{11} M_\odot$ (plain black histograms) and $10^{11}$ -- $10^{12} M_\odot$ (shaded red histograms).  }
\label{fig:sfr_mass}
\end{figure}

A direct relation between the stellar mass and the star formation
rate in high redshift galaxies was identified both at
$z\simeq 1$ \citep{elbaz07} and in a subsample of star-forming BzK--SF at
$z\simeq 2$ \citep{daddi07a}. To some extent, this disagrees  with analogous trends for galaxy
properties in the local Universe, where the most massive
galaxies have very low levels of star formation rate \citep{heavens04}. We illustrate the
global evolution in this relation by plotting in
Fig. \ref{fig:sfr_mass} the star formation rate of each galaxy as a
function of its corresponding stellar mass.

It is immediately clear that a trend between star formation rate and
stellar mass is present at all redshifts. This relation is of course neither
tight nor unique: at all redshifts there exists
a fraction of galaxies with low star formation rates,  as also shown by the inset histograms. The SFR distribution becomes more and more bimodal with decreasing redshift. Figure \ref{fig:sfr_mass} also shows that at the highest redshifts only the most massive galaxies are already experiencing a quiescent phase. 
We note that we are incomplete at low masses and SFRs because of the magnitude limit of the sample. 
Nevertheless, there is a clear trend between SFR and M of the
star-forming galaxies, and to quantify this, we computed a
$2\sigma$--clipped least squares fit, assuming the relation $SFR = A
(M/10^{11}M_\odot)^\beta$. We found values for ($A$, $\beta$) equal to (23.47, 0.70), (38.93, 0.73),
(46.26, 0.65), (127.98, 0.85), respectively in the four redshift bins from $z\sim 0.3$ to $z\sim2.5$. 
The increase in the normalization $A$ is robust, and caused by the global increase in the star formation rate with redshift. The steepening of the slope, instead, is less robust, since it depends on the threshold used for the $\sigma$--clipping and on the SED fitting parameters, e.g., the minimum galaxy age. With a $3\sigma$--clipping, slopes can vary by $\sim 20$\%. 
Considering these uncertainties, our results are in broad agreement with previous studies \citep{daddi07a,elbaz07}. 

It is interesting to note that the correlation between SFR and stellar
mass holds also using the \sfrfits  estimates. While the two
correlations are consistent at intermediate redshifts, the slope of the
\sfrfits one is steeper at low redshift (0.90) and milder at high redshift (0.42), because of the systematic
trends shown in Fig. \ref{fig:sfrratio}. 

\subsection{The evolution in the cosmic star formation rate density} \label{sec:sfrd}

\begin{figure}[!t]
\resizebox{\hsize}{!}{\includegraphics[angle=0]{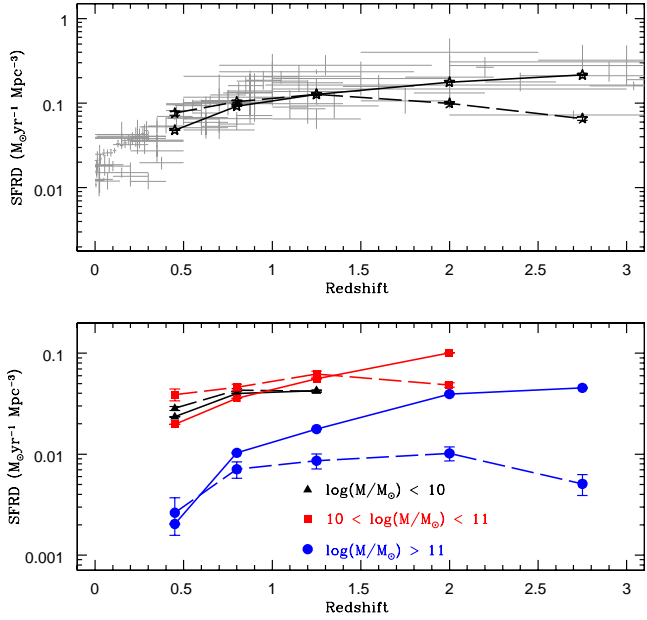}}
\caption{
Upper panel: the redshift evolution of the total cosmic SFRD. Solid lines refer to IR--based estimates, dashed lines to SED fitting estimates. The compilation of \cite{hopkins06} is shown as small gray dots. 
Lower panel: the redshift evolution of the cosmic SFRD by galaxies of different
  masses. Only galaxies above the completeness limit in stellar mass at each redshift (see last column of Table \ref{tab:t1}) are shown. Solid lines refer to IR--based estimates, dashed lines to SED fitting estimates. Black, red and blue symbols correspond to different mass ranges as shown in the
  legend.  Error bars include uncertainties on the SFR estimates and Poissonian errors. 
   }
\label{fig:sfrd}
\end{figure}

The most concise representation of the evolution in the star formation
rate across cosmic time is the star formation rate density (SFRD),
which we show in Fig. \ref{fig:sfrd}. 
We show both the IR--derived estimates (continuous lines) as well as
those obtained purely by the SED fit (dashed line).  
In this figure only, we extend the analysis to the redshift bin $2.5<z<3$.
In this redshift regime, the IR templates are hardly representative, since the \mips 
 band is no longer representative of the dust emission peak, although the \cite{dh02} models are still formally applicable. For these reasons, the \sfrirs estimate is  tentative and should be interpreted with caution.
 
Because of the complex selection criteria that we adopted, it was not easy to compute the corrections necessary to include
 the contribution of galaxies fainter than those included in our
 sample.  Because of our mid-IR selection, we assumed that we include nearly all highly star-forming galaxies. The missed fraction consists therefore mainly of galaxies with low star formation activity, sampled by the $z$--selection. A good approximation of our incompleteness can be obtained by fitting the star formation rate function computed on the $z$--selected sample in each redshift bin. Volumes were computed with the 1/V$_{max}$ method. We found that the SFR functions can  be fitted by Schechter functions of slopes $\sim -1.5$. We then corrected our observations for these contributions at low SFR values by extrapolating from the fitted functions.  The corrections are small for $z<2.5$, of the order of 10\%, consistent with the assumption that the sample is complete at high star formation rate values. At $z>2.5$, the corrections are higher than 50\%.
 
We first note that the total SFRD (upper panel of Fig. \ref{fig:sfrd}) which we derived from our sample (black stars) reproduces well the
compilation of other surveys completed by \citet{hopkins06}, normalized to
a standard \cite{salpeter55} IMF, and is also in agreement with \cite{lefloch05}, \cite{caputi07} and \cite{rodighiero09}. 

It is more interesting to study the trend in the SFRD for galaxies of
different stellar masses (lower panel of Fig. \ref{fig:sfrd}). 
Previous surveys, such as  Combo--17 \citep[][that traced the evolution in the SFRD to $z\simeq 1$]{zheng07} and the GDDS one \citep[][which first presented the evolution in the SFRD in a $K_s$--selected sample to $z\simeq 2$]{juneau05}, demonstrated that the contribution to the SFRD by the more massive galaxies (typically those with $M>10^{11}M_\odot$) is negligible at low redshift, and becomes much larger at $z>>1$ corresponding to the major epoch of formation in the more massive galaxies. 
Our data confirm this picture, with an increase  by a factor 20 from $z\simeq 0.5$ to $z\simeq 2$.   The sharp decline at low $z$ suggests that these structures must have formed their stars at earlier epochs. 
Keeping in mind the caveat mentioned above about the reliability of \sfrirs at $z>2.5$, we note that the increase in the star formation rate density in $M>10^{11}M_\odot$ galaxies appears to halt at $z > 2.5$. This is expected since, at these $z$, the number density of $M>10^{11}M_\odot$ galaxies drops quickly \citep{fontana06,marchesini08}.

The evolution in the SFRD for galaxies of different stellar masses is one of many pieces of evidence of \textit{downsizing}. 
Indeed, \textit{downsizing} is evident as an evolution in the slope of the SFRD as a function of redshift, which becomes yet steeper as high mass galaxies are considered because of the more rapid evolution of massive  galaxies, wherever a number of physical processes are expected to suppress the star formation more efficiently than in lower mass galaxies. 

The exact values of the SFRD depend very sensitively on the mass ranges adopted. Our mass bin choice is determined by the intention of having good statistics in each bin. 
Keeping in mind this warning, our observations seem to support the \textit{downsizing} picture for the reasons mentioned above.   Indeed, in the redshift interval 0.3 -- 1.5, where all mass samples are complete, the SFRD derived from the IR emission increases by a factor 1.8, 2.75, and 9, respectively for bins of increasing mass.  
We also detect a similar steepening in the slope of the relation between the average star formation rate per unit mass (which is treated in Sect. \ref{sec:ssfr}) and redshift between low and high mass galaxies.

\begin{figure}[!t]
\resizebox{\hsize}{!}{\includegraphics[angle=0]{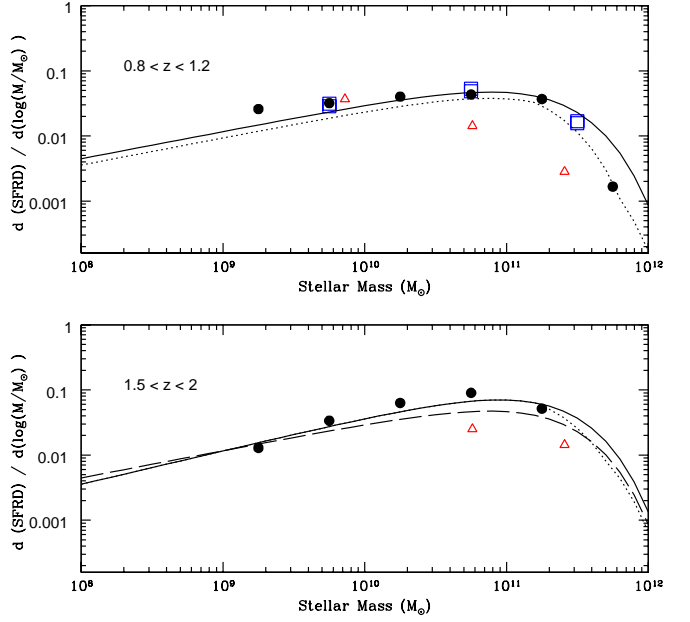}}
\caption{The contribution to the cosmic SFRD by galaxies of given
  mass, in two redshift bins. Solid and dashed lines are computed assuming $f(M)=1$. The dashed line in the lower panel correspond to the $z\simeq 1$ line of the upper panel. Dotted lines are computed assuming $f(M) \neq 1$ (see text). Black filled dots:
  this work; blue squares: Combo-17 \citep{zheng07}; red open triangles: GDDS \citep{juneau05}. }
\label{fig:sfrd_contrib} 
\end{figure}

At all observed redshifts, the more massive galaxies do
not dominate the global SFRD.  To some extent, this may follow from the exact choice of the mass bins. However,  it is easy to predict that the major contributors to the
global SFRD are galaxies immediately below the characteristic mass
$M^*$. 
The number of galaxies of given mass $N(M)$ at any
redshift can be represented as a Schechter function ($N(M) \propto
(M/M^*)^\alpha exp(-M/M^*)$). Given the tight correlation between stellar
mass and star formation rate described above ($\dot{M} \propto
M^\beta$), and assuming that the fraction of star-forming galaxies is
$f(M)$, the contribution to the cosmic SFRD for galaxies in a
logarithmic interval $dlogM$ is simply 
\be
\frac{d (SFRD(M))}{dlog(M)} \propto(M/M^*)^{\alpha+\beta+1} exp(-M/M^*) f(M)
\ee

Since $\alpha$ is in the range $\simeq -1.2 / -1.4$ \citep{fontana06} and $\beta
\simeq 0.6-0.9$ (see above, relating to the IR--based estimate), and ignoring for the moment $f(M)$, the
shape of the $d (SFRD(M)) / dlog(M) $ is that of a Schechter function
with positive slope, which has a peak around $M^*$ and decreases at
$log(M)<log(M^*)$. The shape of $f(M)$ is somewhat more uncertain. 
We simply assume that the fraction of active galaxies above the characteristic mass is around  0.5 at $z\simeq 1$, around 0.8 at $z\simeq 2$, and  will probably be higher at lower masses (around 0.8 at $z\simeq 1$ and around unity at $z\simeq 2$). As a result, it  further decreases the distribution of $d (SFRD(M)) / dlog(M) $ at high masses, without affecting the low
mass regime.

Using the parameters of the galaxy mass function obtained in \cite{fontana06}, and
the SFR--stellar mass correlation shown above, we  computed the expected $d
(SFRD(M)) / dlog(M) $ distribution in two redshift bins, 0.8--1.2 and 1.5--2, where
 data from different surveys \citep{zheng07,juneau05} could be combined. 
To convert masses and SFRs of  \cite{zheng07} from a \cite{chabrier03} to a \cite{salpeter55} IMF, we used a factor of 1.78 \citep{bundy06} and 1.5 \citep{ferreras05}, respectively. Masses and SFRs of \cite{juneau05} were instead renormalized from  a \cite{baldry03} to a \cite{salpeter55} IMF using the factors 1.82 \citep{juneau05} and 2 \citep{hopkins06}.  
We note that the argument presented above is unaffected by the well known discrepancy between the directly measured mass density and the one inferred from the integration of the star formation rate density \citep{wilkins08}. Our point here does not concern the integrated amount of stars formed, but only the analytical shape of the SFRD. 

The result is shown in Fig. \ref{fig:sfrd_contrib}, where we plot the expected $d (SFRD(M)) /
dlog(M) $ distribution by assuming both $f(M)=1$ and a $f(M)$ that decreases with both increasing mass and decreasing redshift. 
Our data are in good agreement with those provided by \cite{zheng07}, who inferred SFR estimates using the same method that we adopt to compute \sfrir. In contrast, \cite{juneau05} estimated the SFR from the rest-frame UV continuum, and obtained values lower  than \sfrir. 
The differences can probably be attributed to the different selection criteria and SFR estimators used in their work.

This analysis confirms that the major contribution to the star formation is from galaxies of masses around or immediately below the characteristic mass $M^*$. This statement is consistent with the evolution in the SFRD for different mass bins that we show in Fig. \ref{fig:sfrd}. The combination of the steepening in the SFR--M relation, the increase in the SFR values with increasing redshift, and the lower fraction of active galaxies at low $z$, is more significant than the effect of the decrease in the characteristic stellar mass at increasing $z$.

\subsection{The specific star formation rate and the comparison with theoretical models } \label{sec:ssfr}

\begin{figure*}[!t]
\resizebox{\hsize}{!}{\includegraphics[angle=0]{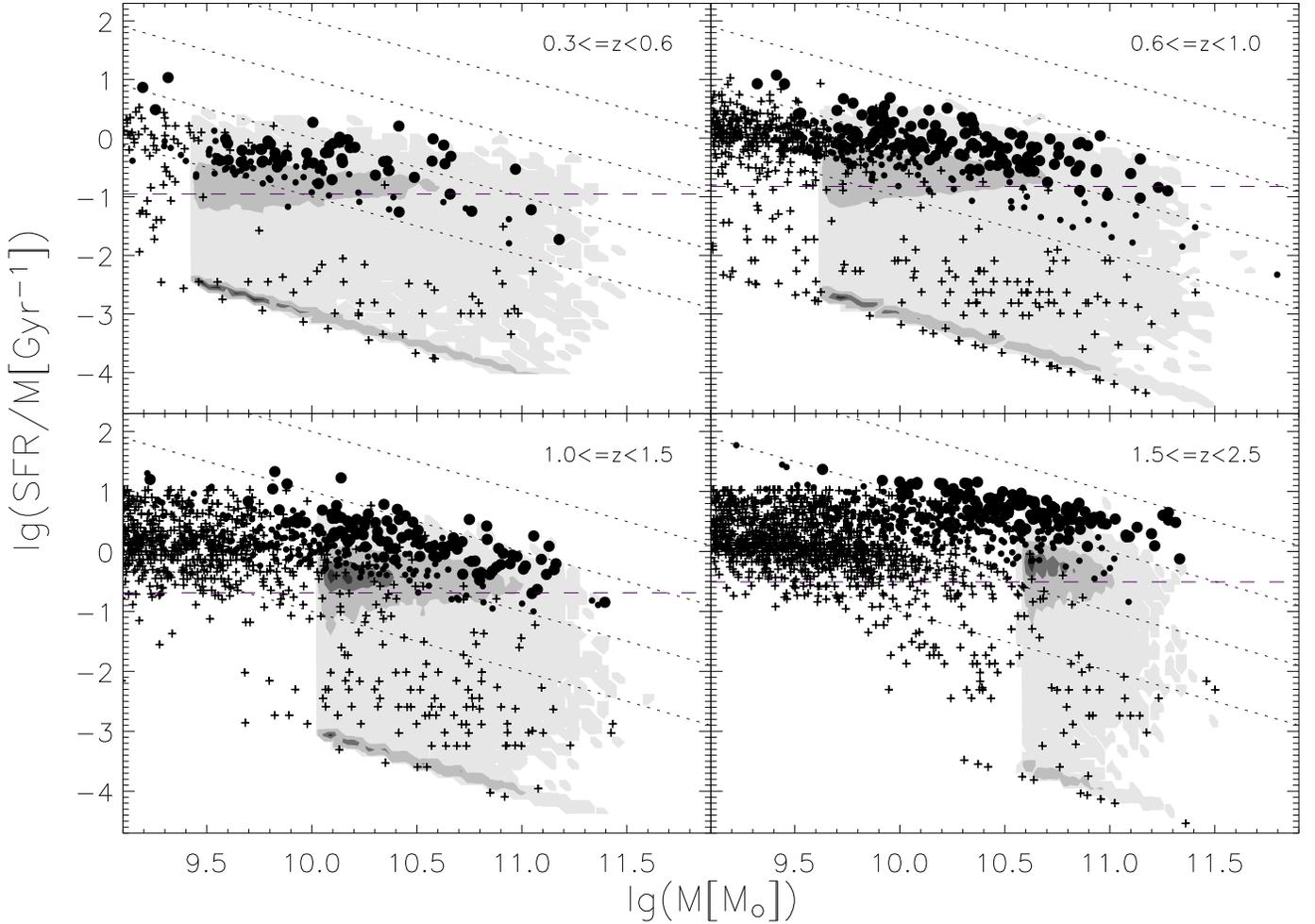}}
\caption{Relation between the specific star formation rate and the stellar mass calibrated to a \cite{chabrier03} IMF. 
Larger dots correspond to \mips sources with F$_{24\mu m} > 83 \mu$Jy, while smaller ones 
show  \mips detections with $20 \mu$Jy $<$ F$_{24\mu m} \le 83 \mu$Jy. Plus refer to \mips upper limits. Dotted lines correspond to constant SFRs of 1, 10, 100 and 1000 $M_\odot yr^{-1}$.  The horizontal dashed lines indicate the inverse of the age of the Universe at the centre of each redshift interval. 
Shaded contour levels (at 0.05\%, 10\%, 50\% and 80\% level) represent the predictions of the \cite{kitzbichler07} rendition of the Millennium Simulation. 
}
\label{fig:mdot}
\end{figure*}

In Fig. \ref{fig:mdot}, we plot the relation between the stellar mass
and the specific star formation rate (SSFR hereafter) for all galaxies
divided into redshift bins. To be able to compare  our findings with the Millennium Simulation predictions, we converted our masses and SFR to the \cite{chabrier03} IMF used by the Millennium Simulation.

We recall that our SSFR estimates are quite low compared to the possible estimates made by adopting standard values for \sfrir. On the one hand, the conversion
between mid-IR flux and SFR includes a lowering factor for high SFRs,
as described in Sect. \ref{sec:sfrestimate} and in Appendix A. On the other side, the use of more recent models
of stellar populations \citep{maraston05} would slightly decrease the
average stellar masses \citep[by 20\% on average, see][]{salimbeni09}, and hence increase the
derived SSFR.

First of all, we notice a strong bimodality in the SSFR distribution. 
Two distinct populations, together with some sources lying
between the two, are detectable, one of young, active and blue
galaxies  (the so-called blue cloud) and the other one consisting of old, ``red and dead'', early-type galaxies  (red sequence) (see also Fig. \ref{fig:sfr_mass}). The loci of these two populations are consistent with
the selection in \cite{salimbeni08} between early- and late-type
galaxies.  It is noticeable how these evolved objects are already in
place even at the highest redshift. We reserve a more complete
discussion of these objects to \cite{fontana09}.

A trend for the specific star formation rate to increase with redshift
at a given stellar mass is evident: galaxies tend to form
their stars more actively at higher redshifts. The bulk of
active sources shifts to higher values of SSFR with increasing
redshift. 
Our findings are in good agreement with
\cite{perezgonzalez05} and \cite{papovich06}.

A significant fraction of the sample, increasing with redshift, is in
an active phase. 
It is natural to compare the SSFR (which has units of the
inverse of a timescale) with the inverse of the age of the Universe at the
corresponding redshift $t_U(z)$. We  define galaxies with
$M/SFR<t_U(z)$ as ``active'' in the following, since they are
experiencing a major episode of star formation, potentially building
up a substantial fraction of their stellar mass in this episode
\footnote{ Indeed, if $M = \left\langle SFR \right\rangle_{past}
  \times t_U(z)$, where $\left\langle SFR \right\rangle_{past}$ is the
  star formation rate averaged over the whole age of the Universe at
  the corresponding $z$, the requirement $SFR/M > (t_U(z))^{-1}$
  implies $SFR > \left\langle
  SFR \right\rangle_{past}$.}. Galaxies selected
following this criterium are forming stars more actively than in their recent past.

\begin{table}
\centering
\begin{tabular} {cccc}
\hline \hline 
\noalign{\smallskip} 
\multicolumn{4}{c}{$\langle$ SSFR [Gyr$^{-1}$] $\rangle$}\\
\hline
$\Delta z$ & IR+UV & SED fitting & M$_{lim}$[$M_\odot$] \\
\noalign{\smallskip} \hline \noalign{\smallskip}
0.3 - 0.6 & 0.278 $\pm$ 0.022 & 0.570 $\pm$ 0.137  & $5\cdot 10^9$\\
0.6 - 1.0 & 0.487 $\pm$ 0.017 & 0.658 $\pm$ 0.103 & $ 8\cdot 10^9$\\
1.0 - 1.5 & 0.754 $\pm$ 0.029 & 0.854 $\pm$ 0.168 & $ 2\cdot 10^{10}$\\
1.5 - 2.5 & 1.652 $\pm$ 0.058 & 0.420 $\pm$ 0.122 & $ 7\cdot 10^{10}$\\
\noalign{\smallskip} \hline \noalign{\smallskip}
\end{tabular}

\caption{Average observed IR-- and fit--based SSFR at different redshifts for galaxies above  
 the completeness limit in stellar mass. 
}
\label{tab:t1} 
\end{table}

At $1.5 \leq z < 2.5$, the fraction of active galaxies in the total sample is
66\%, and their mean SFR is $309\ M_\odot yr^{-1}$. To compute the total stellar mass produced within this redshift interval, it is necessary to know the duration of the active phase. For this purpose, we use a duty cycle argument and suppose that the active fraction of galaxies is indicative of the time interval spent in an active phase. 
We adopt the assumption that the active fraction is stable within the redshift bin considered. The time spanned in the 1.5--2.5 redshift interval corresponds to 1.5 Gyr. By multiplying the fraction of active galaxies by the time available, we derived an average duration of the active phase of 0.99 Gyr. The average amount of stellar mass assembled within each galaxy during these bursts is measured to be the product of the average SFR and the average duration of the active phase, and  equals to $3.1 \times 10^{11}M_\odot$,  representing a significant fraction of the final stellar mass of the galaxies considered.   
Although quite simplified, this analysis implies that most of the stellar mass
of massive galaxies is assembled during a long-lasting active phase at
$1.5 \leq z < 2.5$. It is important to remark that this process of
intense star formation occurs directly within already massive
galaxies, and,  given its intensity, prevails throughout growth episodes due to merging events of already
formed progenitors. 
A similar point is also stated by \cite{daddi07a}.  Independent arguments converging on the same result are based on the tightness of the SFR-mass relation \citep{noeske07}, on the kinematics of disks \citep{cresci09} and on the analysis of the accretion histories of dark matter haloes in the Millennium Simulation \citep{genel08}.

To provide a further physical insight in this process, we 
compared our results with the predictions of three recent theoretical
models of galaxy formation and evolution. Our sample is affected by mass incompleteness, so only galaxies above the completeness limit in each redshift bin were considered in the comparison.  We note that this limit depends on the redshift bin.
In Table \ref{tab:t1}, we report the average SSFR as a function of the
redshift bin for the IR-- and the fit--based estimates, along with the completeness mass cuts computed as described in \cite{fontana06}. 
SSFRs and masses are calibrated to a \cite{salpeter55} IMF.

In Fig. \ref{fig:mdot}, we show the predictions of a semi-analytical rendition of \cite{kitzbichler07} of the
Millennium N-body dark matter Simulation
\citep{springel05,lemson06,delucia07}, which adopts a WMAP1 cosmology. 
We find that the model  predicts an overall
trend that is consistent with our findings. The SSFR
decreases with stellar mass (at given redshift) and increases with
redshift (at given stellar mass). In addition, it forecasts the
existence of quiescent galaxies even at $z > 1.5$.  
However, the average observed SSFR is systematically under-predicted
(at least above our mass limit) by a factor $\sim$3-5 by the
Millennium Simulation. 
A similar trend for the Millennium Simulation at $z\sim2$ 
was already shown by \cite{daddi07a}. 

We also compared our findings with the semi-analytical models of \cite{menci06} (adopting a cosmology consistent with WMAP1) and MORGANA \citep[][updated by \cite{lofaro09}, adopting a WMAP3 cosmology]{monaco07}. They show very similar trends with respect to the Millennium Simulation, with only slightly different normalizations. 

\begin{figure}[!t]
\resizebox{\hsize}{!}{\includegraphics[angle=0]{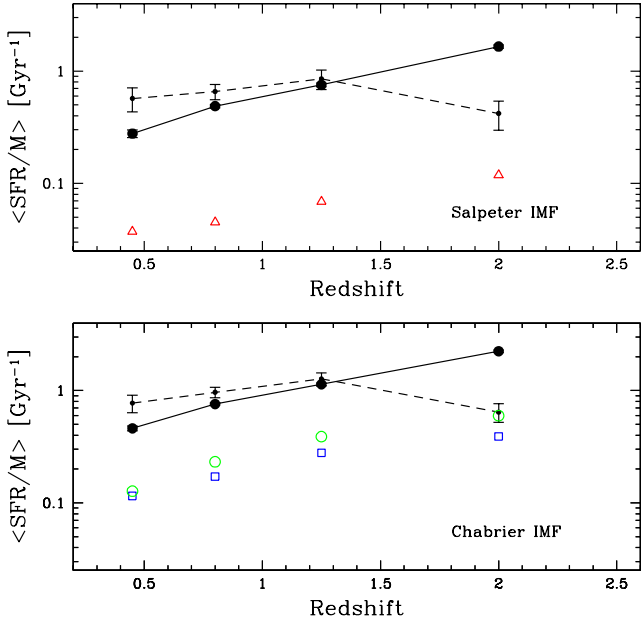}}
\caption{
Average SSFR in each redshift bin. Black solid lines refer to IR--based estimates, dashed lines to the SED fitting estimate.  Error bars for the fit--based estimate are larger than those of the IR--based one because of the intrinsic parameter degeneracy involved with the SED fitting procedure. 
In the upper panel, we show the comparison with \cite{menci06} semi-analytical model (red triangles), which adopts a \cite{salpeter55} IMF. In the lower panel we compare our data with the predictions of the Millennium Simulation \citep{kitzbichler07} (blue squares) and MORGANA (green circles), which adopt a \cite{chabrier03} IMF. 
The comparison is done above the completeness limit in stellar mass, which is different in each redshift bin (see text and Table \ref{tab:t1}).
}
\label{fig:downsizing}
\end{figure}

In Fig. \ref{fig:downsizing}, we plot the average SSFR as a function of the
redshift bin for the IR-- and the fit--based estimates as well as those predicted by the three theoretical models.  Once again, we only consider galaxies above the completeness limit in mass for each 
redshift bin (see last column in Table \ref{tab:t1}). The trend depicted must therefore not be considered  an intrinsic trend, since we  observe different populations at the different redshifts because of the mass cuts.  
Errors in the average SSFR were estimated by a Monte Carlo simulation.

We find that although all models taken into account reproduce the global observed trend, they predict an average star formation activity that is lower than observed in most of the mass regimes. 
The observed star formation occurring in situ inside massive galaxies is higher than predicted by a factor 3--5 for the Millennium Simulation and for MORGANA, and around 10 for \cite{menci06} model. 
 \cite{daddi07a} claimed that the star formation rates predicted by the Millennium Simulation are as much as one order of magnitude lower than those observed at $z\sim 2$.  Similar mismatch at these redshifts between observed SFRs and those expected by various kind of theoretical models, independent of their
physical process implementations, were found and discussed by \cite{dave08}. To reconcile data and model predictions, \cite{khochfar08} proposed a galaxy formation scenario at high redshift that is mainly driven by cold accretion flows. Their model allows an increased star formation efficiency  resulting in a closer agreement with observations.
 
\begin{figure}[!t]
\resizebox{\hsize}{!}{\includegraphics[angle=0]{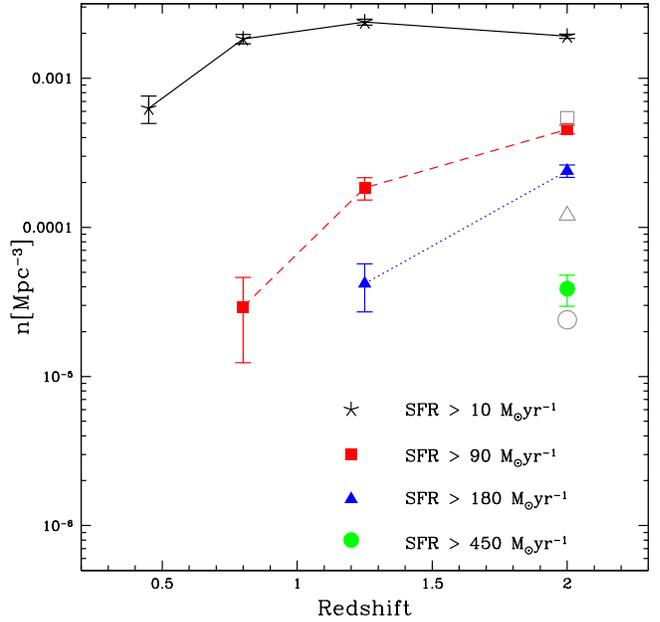}}
\caption{  Number density of star-forming galaxies as a function of redshift and lower SFR. Filled symbols represent our observations corresponding to the SFR thresholds shown in the legend. Open gray symbols at $z \sim 2$ are the predictions of \cite{khochfar08b} model. }
\label{fig:nd}
\end{figure}

We then considered the number density of galaxies with star formation rates higher than a fixed threshold as a function of redshift, and compared our observations with the model of \cite{khochfar08b} at $z\sim 2$. Figure \ref{fig:nd} compares our data (filled symbols) with the  theoretical predictions (open symbols). The SFR limits were chosen  to allow  comparison with  \cite{khochfar08b} results\footnote{Note that they adopt a \cite{chabrier03} IMF, so the SFR thresholds have been renormalized by adopting the conversion factors in Sect. \ref{sec:sfrd}. }. The agreement is very good for objects with $SFR>90 M_\odot yr^{-1}$, while the model slightly underpredicts the number density of galaxies with higher levels of star formation rate.

A more comprehensive  comparison between theoretical predictions and observations was presented in \cite{fontanot09}.

\section{Summary and discussion}

We have presented  a revised version of our GOODS-MUSIC
photometric catalog for the GOODS-S field. The major new feature of
this release, on which our scientific discussion is based, is the
inclusion of \mips data taken from the Spitzer MIPS public images,
of which we provide a self-consistent photometry for each object in
the catalog. We employed a PSF-matching technique, performed by ConvPhot software
\citep{desantis07}, to measure \mips photometry by exploiting the high
resolution $z$-ACS image used as a {\it prior} to detect the objects'
positions. This allows us to reduce the effects of confusion noise and
deblending problems caused by the large size of the MIPS PSF.

We have used this new catalog to study the star-forming properties of
galaxies up to $z\simeq 2.5$. 
We have first compared the estimates of the
SFR obtained from the total IR luminosity  (\sfrir) and from the SED fitting
analysis to the overall 0.3 to 8 $\mu$m photometry (\sfrfit), which are the two
major estimators of the SFR that have been used so far in high redshift galaxies. 
We found that the two tracers are  consistent overall, especially in the
redshift range 0.3 -- 1.5. The overall median ratio of \sfrirs to
\sfrfits is around unity, a limited number of objects appearing to have 
discrepant results. The agreement between the two estimators appears
to depend on redshift because the scatter increases with redshift. 
Moreover, \sfrfits is slightly systematically overestimated with respect to
\sfrirs in the lowest redshift bin ($0.3-0.6$). The scatter increases
significantly at $z>1.5$, where the IR+UV value is systematically
larger than the one resulting from the SED fitting analysis for bright
objects.  However, these results stem from a systematic trend that
holds at all redshifts. In galaxies with star formation rates
$<10M_\odot/yr$, the fit--derived SFR is on average higher than \sfrir,
while the opposite holds at SFR $>100M_\odot/yr$. 
It is at present difficult to ascertain the origin of this systematic
trend.  It can be due to systematics that affect the interpretation of the data, either
from the SED fitting technique as the target galaxies move in
redshift, or in the templates used in the extrapolation of the mid-IR
observed flux, especially at high redshift, where it samples the PAH
region. Alternatively, it could have physical origins, such as a
metallicity trend or a failure in the assumption that a single
attenuation law can adequately model the output from a star-forming
galaxy.

Keeping in mind these uncertainties in the estimate of the
star formation rate, we summarize  the basic results obtained by adopting \sfrir, which we assume to be a more reliable tracer because it is not affected by dust extinction.

\begin{itemize}

\item  We show that, at all redshifts considered here, there is a
correlation between the stellar mass and the star formation rate of
star-forming galaxies. The logarithmic slope of this correlation,
after applying a $\sigma$--clipping algorithm to remove all quiescent galaxies, is
in the range $0.6-0.9$, with some indications of a steepening with increasing redshift.

\item The SFRD derived from our sample agrees with the global trend
already depicted by other surveys. When divided according to stellar
mass, it shows that more massive galaxies enter  their active phase
at redshifts higher than lower mass ones. At $z>2.5$, the increase in
the SFRD due to the more massive galaxies  (with $M>10^{11}M_\odot$) appears to halt, in broad
agreement with the expectations of theoretical models
\citep{menci04,bower06}. This is mainly because galaxies of higher mass become extremely rare at these redshifts. The increase in the slope of the SFRD with redshift of samples of increasing mass seems to support the {\it downsizing} scenario.

\item At all redshifts, the main contributors to the cosmic SFRD are
galaxies around, or slightly below, the characteristic stellar mass
$M^*$. 

\item Massive galaxies at $z\simeq 2$ are
vigorously forming stars, typically at a rate of $300 M_\odot$yr$^{-1}$. A simple duty--cycle argument  (see Sect. \ref{sec:ssfr}) suggests that they
assemble a significant fraction of the final stellar mass during this
phase.

\item The specific star formation rate of our sample shows a well-defined bimodal
distribution, with a clear separation between actively star-forming
and passively evolving galaxies. 

\end{itemize}

While these results are grossly independent of the particular star formation rate estimate, the specific details appear to depend on the
chosen indicator. In particular, the correlation between the SFR and
the stellar mass still holds using \sfrfit, but it is steeper at $z\simeq
0.5$ and it becomes flatter at high redshift, remaining similar in the
two intermediate redshift bins. 
As far as the star formation rate density and the specific star
formation rate are concerned, the two techniques used to estimate the
SFR provide consistent results up to $z\sim 1.5$. However, significant
differences are evident at redshift $\gtrsim 1.5$, where the IR--based SFRD
flattens and the fit--based one begins to decline. The average
IR--based SSFR  increases monotonically to the highest observed
redshifts, while the fit--based one has a turn-over around $z\sim1.5$
and then decreases. The trends above directly reflect the correlation
between \sfrirs and \sfrfit.

We used our results for the redshift evolution in the specific star
formation rate, and its trend with the stellar mass, to investigate
the predictions of a set of theoretical models of galaxy formation in
$\Lambda$-CDM scenario.

On the one hand, these models reproduce the global trend that we find
in the data -- the most important being the increase in the specific
star formation rate with redshift, and its trend with stellar
mass. Somewhat surprisingly, however, the average SSFR of galaxies in
our sample is significantly higher than predicted by theoretical
models, in most of the mass regimes. Essentially, after including a
strict completeness limit in stellar mass, we found that the typical
SSFR of galaxies at a given mass is a factor at least $\simeq 3-5$ higher than
predicted by the models we have considered here. This mismatch is very
clear for massive galaxies ($M_*\simeq 10^{11} M_\odot$) at $z\simeq 2$ and for
less massive galaxies at $z\simeq 1$, where we are able to complete a higher quality 
statistical analysis.

It is not obvious to ascertain the origin of this mismatch.  It could be caused by a genuine failure of the models.
These models often quench the star formation in
massive galaxies to prevent the formation of blue, giant galaxies at
low redshift, and to reproduce the existence of red, massive galaxies
at high redshift. The mismatch that we observe could imply that the
feedback and star formation recipes adopted are too simplified or
incorrect.  
Most of the models considered in this work adopt the commonly used star formation scenario, where gas uniformly collapsing towards the galaxy centre forms a stable disk. New processes are  being explored \citep[e.g.,][]{dekel09} that involve a more rapid formation of galaxies by cold streams and  could lead to closer agreement with observations.

Alternatively, the mismatch could be related to the overestimate of the
stellar mass of the typical star-forming galaxy -- for instance,  due
to a combination of the overestimate of merging events and of the star formation activity in its past history, i.e., at redshifts higher than those sampled in this work. A possible consequence of this is the overestimate of the stellar mass function predicted by the Millennium Simulation at high redshifts in relation to  observations \citep{kitzbichler07}.

Additional evidence of an incorrect treatment of the star formation processes was presented by \cite{fontanot09} and \cite{lofaro09}. 
In their models, these authors found an excess of low mass galaxies at $z<2$ and faint LBGs at $z>3$, respectively, which is probably counterbalanced by the suppression of the star formation to reproduce the observed evolution in the SFRD. 

However, we  remark that the interpretation of the observations is
affected by a number of uncertainties, such as uncertainties in the
stellar mass estimates or uncertainties originating  from both the templates
used to convert \mips fluxes into total infrared luminosities (see
Appendix A) and  the SED fitting analysis.  An example of these uncertainties is illustrated by the mismatch
between the integrated star formation rate density and the stellar
mass density. 
This disagreement could be alleviated by an evolving IMF \citep{wilkins08,dave08}, which would provide lower values of SFR.

To make conclusive, quantitative statements in this direction, we
ultimately need to improve the reliability of the SFR measurements,
especially for high redshift galaxies. Forthcoming IR facilities, such as Herschel and ALMA,
will probably provide us with a more coherent picture.

\begin{acknowledgements}
We are grateful to Pierluigi Monaco, Niv Drory and Andrew Hopkins for the useful discussions. 
Observations were carried out with the Very Large
Telescope at the ESO Paranal Observatory under Program IDs LP168.A-0485 and
ID 170.A-0788 and the ESO Science Archive under Program IDs 64.O-0643,
66.A-0572, 68.A-0544, 164.O-0561, 163.N-0210, and 60.A-9120. The Millennium Simulation databases used in this paper and the web application providing online access to them were constructed as part of the activities of the German Astrophysical Virtual Observatory. This work has been partly funded by ASI, under the COFIS contract.
\end{acknowledgements}

\bibliographystyle{aa}

\begin{appendix} 
\section{The estimate of total infrared luminosity} 

We describe our method to convert mid-IR fluxes into total infrared luminosities and compare the results obtained from different template libraries.

By assuming that the IR emission is primarily caused by dust heating during star formation, the SFR remains proportional to the dust thermal emission of the galaxy. \cite{kennicutt98} asserted that the total infrared luminosity \lirs emitted between 8 and 1000 $\mu$m is a good tracer of the SFR. 
At the redshift of our interest ($z \sim 0.3 - 2.5$), MIPS \mips band probes the rest-frame mid-IR emission, which has been demonstrated to correlate with \lirs \citep[e.g., ][]{ce01,elbaz02,papovich02,takeuchi05,zhu08}.

Since dust emits most of its light at longer wavelengths, 
\lirs extrapolation from \mips observations can be a difficult procedure. The rest-frame region under study is particularly complex
given the presence of polycyclic aromatic hydrocarbon (PAH) lines. In principle, these aspects could easily lead to non-negligible errors in the determination of \lirs values. 
Therefore, we employed two different methods to infer \lir, examining and comparing both synthetic and empirical templates using our $K_s$--selected subsample (subsample A). 

\begin{figure}[!t]
\resizebox{\hsize}{!}{\includegraphics{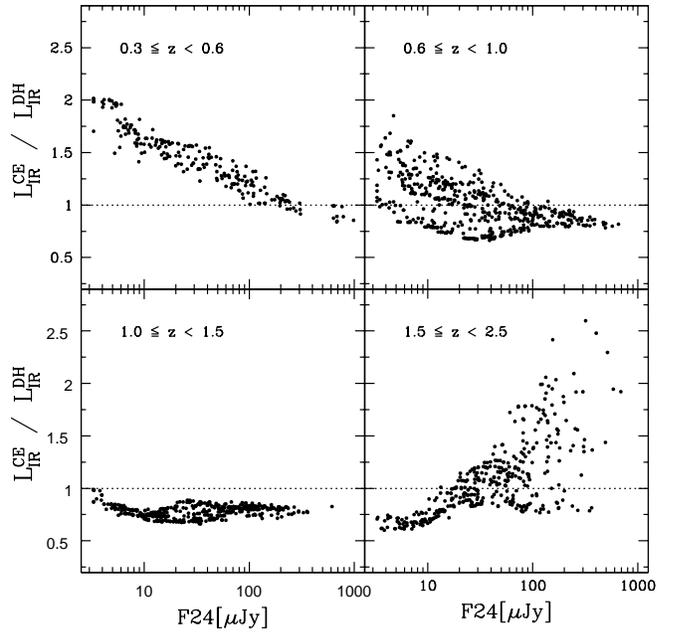}}
\caption{Comparison between the total infrared luminosity estimated with \cite{dh02} and \cite{ce01} templates in different redshift bins. 
  The two synthetic libraries predictions are in good agreement within
  a factor $<3$.}
\label{fig:lirsint}
\end{figure}

Firstly, according to procedures normally adopted in the literature, we considered \cite{dh02} (DH hereafter) and \cite{ce01} (CE hereafter) synthetic libraries. These models do not extend to sufficiently short wavelengths to fit the SED shape, i.e., they do not include the stellar contribution, and at $z \gtrsim 1$ only MIPS \mips band data can be fitted, the other bands moving out of the model range or being shifted to a wavelength region dominated by star light. 
Each model is associated with a given \lir.
In the case of the CE library, each model is provided with its absolute normalization, and hence with a given total infrared luminosity. 
As for the DH library, we assigned a given \lirs to each template using the empirical relation in \cite{marcillac06} between $L_{IR}$ and the predicted $f^{60}_\nu/f^{100}_\nu$ colour. 
In both cases and for each source, once the $k$--correction had been applied, we selected the model that reproduced the \mips observed luminosity and normalized it using the flux difference between the model and the observed galaxy. 
\\
In Fig. \ref{fig:lirsint}, we show a comparison between \lirs  predicted by the two different model libraries. As already noticed in literature \citep{papovich06,marcillac06}, they give consistent results to within a factor 2-3, with the highest differences appearing at high redshifts and high luminosities. However, it is possible to observe a few trends with both luminosity and redshift that depend on the specific template details.

\begin{figure*}[h]
\resizebox{\hsize}{!}{\includegraphics[angle=270]{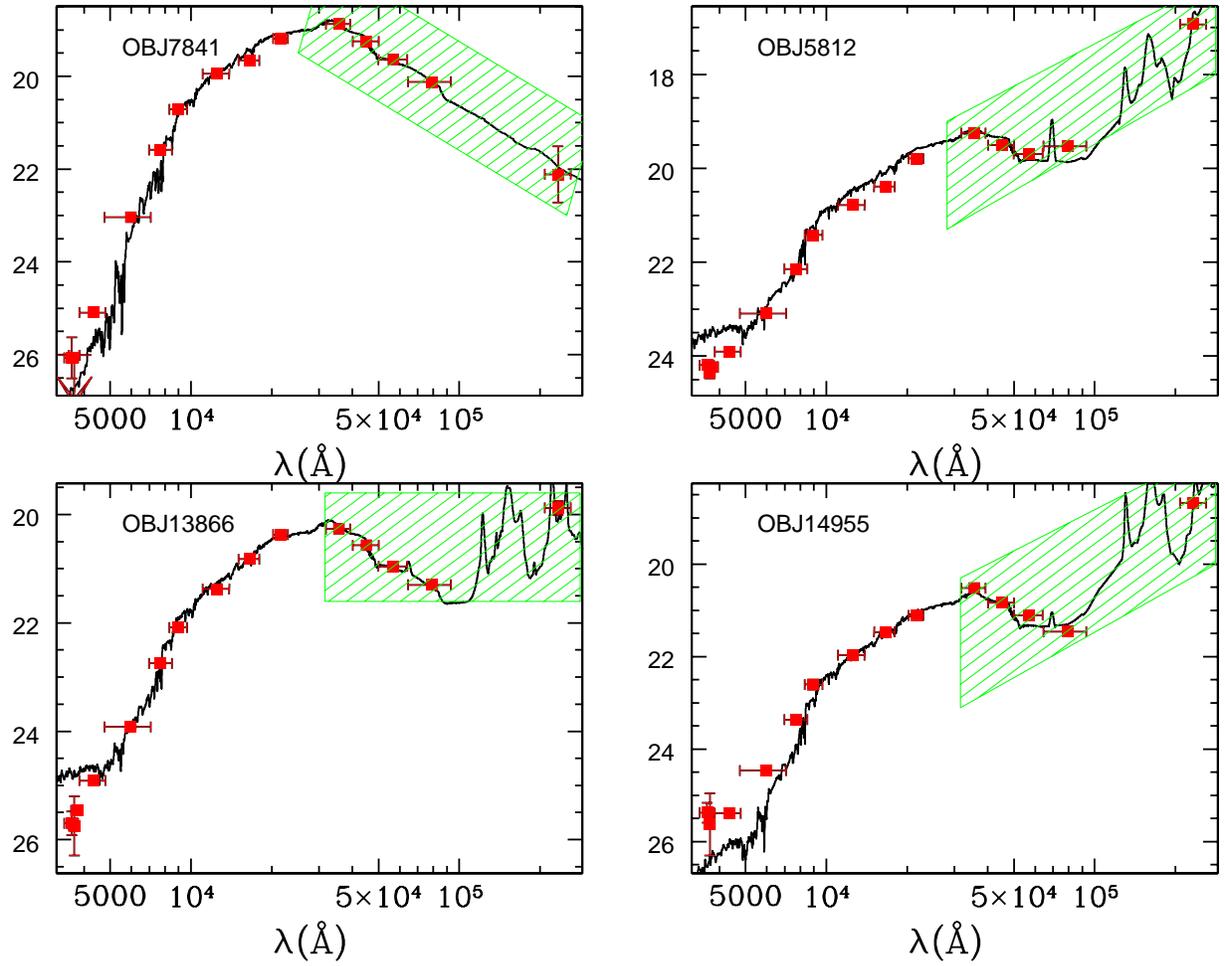}}
\caption{Different typical observed spectral energy distributions fitted by \cite{polletta07} templates. From top left, in anticlockwise direction, galaxies have been fitted by an elliptical, 
a spiral and two starbursts templates. The green shaded region indicates the wavelength range used for the fitting procedure. 
}
\label{fig:sedpolletta}
\end{figure*}

\begin{figure}[!t]
\resizebox{\hsize}{!}{\includegraphics{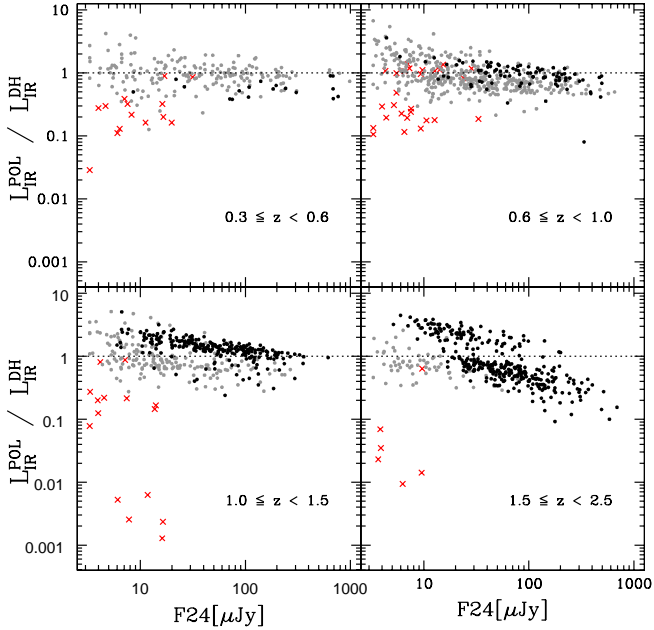}}
\caption{Comparison between the total infrared luminosity estimated with \cite{dh02} and \cite{polletta07} templates. Black and gray dots and red crosses indicate objects fitted respectively by starburst, spiral-like and early-type templates by \cite{polletta07}.   Synthetic models appear to overestimate \lirs at bright luminosities
  in the highest redshift bin when compared to empirical templates.}
\label{fig:lirpol}
\end{figure}

A new kind of approach in our work consists of employing empirical
local spectra to fit the overall galaxy SED shape instead of the \mips
luminosity. These spectra include the stellar contribution
as well as the effects of dust emission.  We considered \cite{polletta07} spectra
(POL hereafter). This library comprises early-type, spiral, starburst,
and different kinds of AGN local templates, ranging from $\sim$100
$\AA$ to $\sim$5 mm. 
The wide wavelength extension allows us to fit the SED shape using the
multiband catalog. After performing several tests, we decided to
carry out a fit over 5 bands (IRAC bands + \mip), excluding the optical range, where the evolution in the stellar component has a more significant effect than the dust emission. 
Since we removed AGN-dominated objects from our
sample, we only fit early-type and star-forming models. If we instead
include AGN SEDs, the estimated \lirs is only poorly affected, resulting
in slightly lower values \citep[see also][]{papovich07}.  Despite the poor
statistical weight (one amongst five bands) and higher noise, the
\mips band is still fitted acceptably. The mean deviation between
observed and fitted magnitudes for \mips detections is $(-0.28 \pm
0.53)$ mag,  68\% of objects being consistent within 0.48 mag and 95\% of
objects being consistent within 1 mag (considering the $F_{24}>20\ \mu$Jy subsample).  An example of different kinds of spectral energy distributions fitted by \cite{polletta07} templates is reported in Fig. \ref{fig:sedpolletta}.
\\
Figure \ref{fig:lirpol} shows the comparison between \lirs estimates
obtained with the DH synthetic library and the POL empirical one. Source
classification based on the spectral shape fitting agrees with that
assumed from \mips emission. As expected, the only galaxies fitted by
early-type SEDs (red crosses) have very faint \mips
emission. Moreover, objects tend to be fitted by starburst-like models
(black dots) rather than spiral-like ones (gray dots) as redshift
increases. Although the scatter between the two \lirs estimates is
larger than in the previous case (note that, in contrast to Fig. \ref{fig:lirsint}, the y-scale is logarithmic), we found global consistency between
the two adopted procedures.  The only relevant deviation affects the
highest redshift bin: in this redshift range synthetic models appear
to overpredict, by up to a factor of 10, \lirs for bright objects. The same
behaviour was observed by \cite{papovich07}, where \mips flux,
fitted with DH library, seems to overestimate the SFR (which is almost
proportional to \lir) with respect to the case where longer
wavelengths (70 and 160 $\mu$m MIPS bands) are considered as
well. They corrected this trend using an empirical second-order
polynomial. A similar trend of the \mips flux overestimating \lirs at bright luminosities
was also pointed out by \cite{bavouzet07} and \cite{rieke09}. 
In our work, we used the \cite{dh02} library for
consistency with previous works and for which the lowering correction needed to estimate the SFR was provided by \cite{papovich07}.

\begin{figure}[h]
\resizebox{\hsize}{!}{\includegraphics[angle=0]{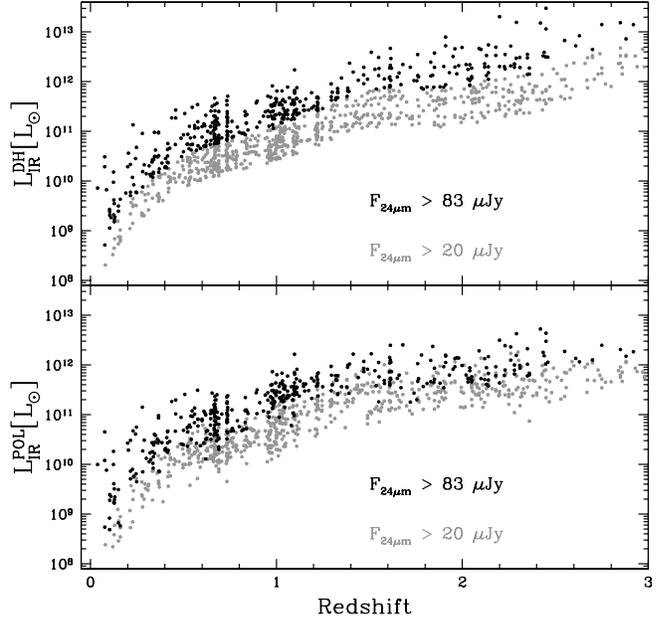}}
\caption{Relation between total infrared luminosity \lir, computed using \cite{dh02} (upper panel) and \cite{polletta07} templates (lower panel), and redshift. Black dots represent fluxes above 83~$\mu$Jy, gray + black dots refer to our sample limited at 20 $\mu$Jy. 
} 
\label{fig:lir}
\end{figure}

In Fig. \ref{fig:lir}, we show \lir, estimated using both synthetic (DH) and empirical (POL)
templates, as a function of redshift. Once
again, the two libraries provide fairly consistent results. All in all,
the conversion between fluxes and total infrared luminosity does not
depend on the assumed templates in a significant way. The observed
\lir--redshift relation  agrees with that presented by
\cite{perezgonzalez05}. As expected, we are only able to detect the
more luminous sources as we move to higher redshift.

\end{appendix}

\begin{appendix}

\section{Error analysis for \sfrfit} 

In Sect. \ref{sec:sfrcomp}, we  compared two different tracers for the star formation rate, one derived from the \mips emission and the other inferred from a SED fitting technique. Although the two estimators were found to be  consistent overall, in a more careful analysis they were found to exhibit a systematic trend depending on the SFR itself as discussed in the text. To assess the reliability of the SFR measurements inferred from the SED fitting, and hence concrete evidence of this characteristic trend, we applied an error analysis similar to that widely adopted in similar cases \citep{papovich01,fontana06,fontana09}. 
Briefly, the full synthetic library
used to identify the best fit spectrum was compared with the observed
SED of each galaxy. For each spectral model (i.e., for each combination
of the free parameters $age$, $\tau$, $Z$, $E(B-V)$), the probability
$P$ of the resulting $\chi^2$ was computed and retained, along with the
associated SFR.  
In the case of galaxies with only photometric $z$, an additional source of error was the redshift uncertainty. To account for this, the error analysis was completed by allowing the redshift to be a free parameter at a local minimum around z$_{phot}$.  

\begin{figure}[!t]
\resizebox{\hsize}{!}{\includegraphics[angle=0]{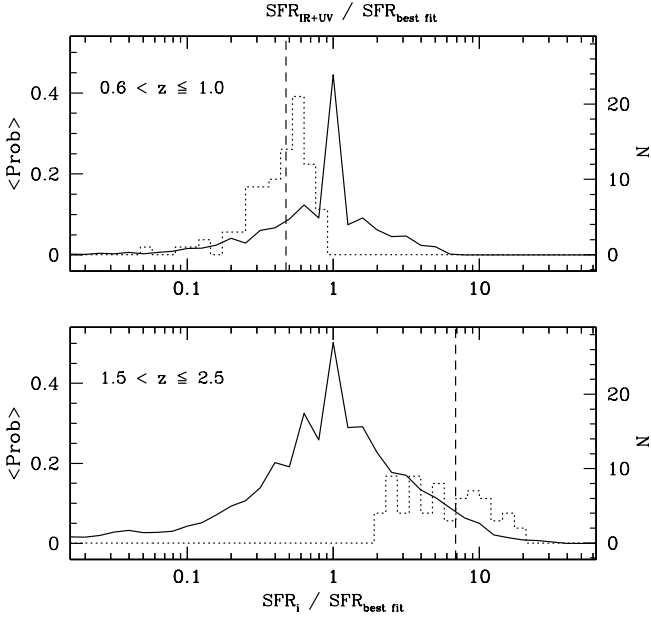}}
\caption{Probability distribution of the star formation rates estimated through the SED fitting technique averaged over each subsample. Upper and lower panels show  subsample A1 and A2 (see text), respectively. The solid line shows the probability associated with the SFR inferred from the generic fitted template, SFR$_i$, as a function of the ratio SFR$_i$/SFR$_{best fit}$. The dotted line shows the distribution of \sfrir/SFR$_{best fit}$ for all the galaxies in the subsample. The dashed line shows the average value for \sfrir/SFR$_{best fit}$. 
} 
\label{fig:chi2}
\end{figure}

We  chose two subsamples by performing the error analysis, for the cases \sfrirs $<<$ \sfrfits and  \sfrirs $>>$ \sfrfit. These two subsamples were called subsample A1 and subsample A2, and  selected to complete a reliable statistical analysis in both cases. Subsample A1 is made of galaxies having $0.6 \leq z < 1.0$, \sfrirs $< 10 M_\odot yr^{-1}$ and \sfrir/\sfrfits $< 0.8$. Subsample A2 consists of galaxies in the redshift bin 1.5 -- 2.5, with \sfrirs $> 200 M_\odot yr^{-1}$ and $2 < $ \sfrir/\sfrfits $< 20$. Highly obscured AGN candidates were removed from both subsamples.

The results of our analysis are shown in Fig. \ref{fig:chi2}. We show the probability distribution of the star formation rates estimated by the SED fitting averaged over each subsample. We plot the probability associated to the SFR inferred from the generic fitted template, SFR$_i$, as a function of the ratio SFR$_i$/SFR$_{best fit}$. We also show the distribution and the average value of \sfrir/SFR$_{best fit}$. 

As expected, the probability curve is much wider at higher redshifts than at lower $z$, where  the spread in our data also becomes larger (see Fig. \ref{fig:sfrratio}). This is caused by the faintness of the galaxies observed at these redshifts and to the larger $k$--corrections. 
However, it is clear from Fig. \ref{fig:chi2} that the SED fitting for galaxies in both subsamples is well constrained, and that the inferred average SFR is significantly higher (lower) than the one derived from the IR emission. Indeed, the average \sfrir/SFR$_{best fit}$  (dashed line) as well as their distribution (dotted line), occupy the tail of the SFR$_i$/SFR$_{best fit}$ distribution. 

We also inspected the individual probability curves, considering, for each source, the ratio of the probability that the SFR from the SED fitting equals  \sfrirs to the best-fit probability  P(SFR$_i$$=$\sfrir)/P(SFR$_i$$=$SFR$_{best fit}$). The fraction of galaxies with a ratio higher than 0.4 is 17.4\%  and 7.2\% in subsamples A1 and A2, respectively. 

Finally, according to a Kolmogorov-Smirnov test, the probability that the SFR values derived from the two tracers are drawn from the same distribution is negligible ($9.5 \times 10^{-15}$ and $7.3 \times 10^{-20}$  for subsample A1 and subsample A2, respectively).

\end{appendix}

\end{document}